\documentclass[]{aa}  
\usepackage{graphicx}
\usepackage{txfonts}
\usepackage{graphicx}
\usepackage[shortlabels]{enumitem}
\usepackage{tabularx}
\usepackage{mathtools}
\usepackage{amsmath}
\usepackage{xcolor}
\usepackage[]{hyperref}
\usepackage{natbib}
\usepackage{url}
\usepackage{times}
\def\spose#1{\hbox to 0pt{#1\hss}}
\newcommand\lsim{\mathrel{\spose{\lower 3pt\hbox{$\mathchar"218$}}
     \raise 2.0pt\hbox{$\mathchar"13C$}}}
\newcommand\gsim{\mathrel{\spose{\lower 3pt\hbox{$\mathchar"218$}}
     \raise 2.0pt\hbox{$\mathchar"13E$}}}

\makeatletter
\renewcommand*\aa@pageof{, page \thepage{} of \pageref*{LastPage}}
\makeatother

\begin{document} 

\title{The disk--torus system in active galactic nuclei: possible evidence of highly spinning black holes}

\titlerunning{Spin and Torus emission in AGNs}
\author{Samuele Campitiello\inst{1,2}\thanks{\email{sam.campitiello@gmail.com}} \and Annalisa Celotti\inst{1,3,4} \and Gabriele Ghisellini\inst{3} \and Tullia Sbarrato\inst{3,5}}

\institute{SISSA, Via Bonomea 265, I–34135, Trieste, Italy \and INAF - Osservatorio Astronomico di Trieste, Via Tiepolo 11, I-34131, Trieste, Italy \and INAF - Osservatorio Astronomico di Brera, Via E. Bianchi 46, I-23807, Merate, Italy \and INFN–Sezione di Trieste, Via Valerio 2, I-34127 Trieste, Italy \and Dipartimento di Fisica "G. Occhialini", Università di Milano - Bicocca, Piazza della Scienza 3, I-20126 Milano, Italy }

\date{Received: 21 June 2021 - Accepted: 20 September 2021}

\abstract{We study the ratio $R$ between the luminosity of the torus and that of the accretion disk, inferred from the relativistic model KERRBB for a sample of approximately $2000$ luminosity-selected radio-quiet Type I active galactic nuclei from the Sloan Digital Sky Survey catalog. We find a mean ratio $R \approx 0.8$ and a considerable number of sources with $R \gsim 1$. Our statistical analysis regarding the distribution of the observed ratios suggests that the largest values might be linked to strong relativistic effects due to a large black hole spin ($a>0.8$), despite the radio-quiet nature of the sources. The mean value of $R$ sets a constraint on the average torus aperture angle (in the range $30^{\circ} < \theta_{\rm T} < 70^{\circ}$) and, for about one-third of the sources, the spin must be $a > 0.7$. Moreover, our results suggest that the strength of the disk radiation (i.e., the Eddington ratio) could shape the torus geometry and the relative luminosity ratio $R$. Given the importance of the involved uncertainties on this statistical investigation, an extensive analysis and discussion have been made to assess the robustness of our results.}

\maketitle

%%%%%%%%%%%%%%%%%%%%%%%%%%%%%%%%%%%%%%%%%%%%%%%%%%%%%%%%%%%%%%%%%%%%%%%%%%%%%%%%
%%%%%%%%%%%%%%%%%%%%%%%%%%%%%%%%%%%%%%%%%%%%%%%%%%%%%%%%%%%%%%%%%%%%%%%%%%%%%%%%
%%%%%%%%%%%%%%%%%%%%%%%%%%%%%%%%%%%%%%%%%%%%%%%%%%%%%%%%%%%%%%%%%%%%%%%%%%%%%%%%
%%%%%%%%%%%%%%%%%%%%%%%%%%%%%%%%%%%%%%%%%%%%%%%%%%%%%%%%%%%%%%%%%%%%%%%%%%%%%%%%

\section{Introduction} \label{sec:intro}

The unification paradigm for active galactic nuclei (AGNs) assumes the presence of dust surrounding the nuclear regions, causing the apparent differences in the observed broad-line emission and X-ray properties (\citealt{Anto}; \citealt{GhiHa}; \citealt{UrPad}).

This dusty and optically thick material would partly "cover" the accretion disk (AD), its corona, and the broad line region (BLR) surrounding the central supermassive black hole (SMBH), absorbing a fraction of the total optical--UV disk luminosity and re-emitting it in the infrared (IR) band (e.g., \citealt{Reesetal}; \citealt{Neugetal}; \citealt{Barva}).

The properties and the geometrical configuration of the dust are still unclear. Several models have been proposed: a smooth or continuous toroidal dust distribution ("torus") was firstly proposed (e.g., \citealt{PierKro}; \citealt{GraDan}; \citealt{Schart}; \citealt{Fritz}). Then, as it was pointed out that such a structure could be unstable, a clumpy distribution was suggested (e.g., \citealt{KroBeg}; \citealt{Tac94}; \citealt{Nenka, Nenkb}; \citealt{HonKish}), which was also supported by observations (\citealt{Risa}; \citealt{Jaffe}; \citealt{Trist}; \citealt{Zhao}).

The features of the torus have recently been studied using large samples of AGNs (e.g., \citealt{Caldero12}; \citealt{MaWa}; \citealt{Hao}; \citealt{Merloni}). One of the simplest approaches to studying the geometry of the dusty gas is to consider its covering factor (i.e., the fraction of sky covered by the torus as seen from the SMBH). This has been estimated from: (1) the numerical ratio of Type 1 to Type 2 AGNs, in carefully selected (unbiased) samples (e.g., \citealt{LawElv}; \citealt{Law}; \citealt{Simp}); (2) detailed physical models of the torus emission (e.g., \citealt{Ezhi}; \citealt{Zhuang}); and (3) the ratio between the bolometric IR and AD luminosities as inferred from the spectral energy distributions (SEDs, e.g., \citealt{Alonso}; \citealt{Caldero12}; \citealt{CastiDe}; \citealt{Toba}). 

More specifically, \citet{Caldero12} obtained information on the torus aperture angle $\theta_{\rm T}$ (as measured from the disk normal) by comparing the ratio between the IR and AD luminosities with that predicted for the AD angular emission pattern by the non-relativistic Shakura-Sunyaev model (hereafter SS; \citealt{SS}). \citet{Caldero12} used a sample of radio-quiet AGNs from the Sloan Digital Sky Survey (SDSS) catalog (\citealt{York}; \citealt{Schnei}; \citealt{Shenetal11}) with redshift in the range $0.56-0.73$. The authors found that the torus reprocesses on average between approximately one-third and one-half of the disk luminosity, corresponding to $\theta_T \sim 40^{\circ}-60^{\circ}$. \citet{Gu} found a similar result for a sample of low-redshift quasars (QSOs), while for high-redshift ones, the author derived a higher covering factor ($\sim 1$). Similar values were found by \citet{CastiDe} for a small sample of bright flat-spectrum radio quasars. The procedure followed by \citet{Caldero12} to constrain $\theta_{\rm T}$ assumes a simple $\cos \theta$ radiation pattern for the AD (as expected for the SS model), however this is possibly too simplistic because both the relativistic effects and the black hole (BH) spin $a$ modify the radiation pattern, especially in the inner region of the disk, where most of the radiation is produced. 

Here, we aim to infer information about the torus geometry (i.e., its covering factor and/or aperture angle), and possibly to constrain the BH spin by comparing the ratio between the torus and disk luminosities, as inferred from the SEDs, with the predictions for a thin AD around a Kerr BH, as described by the KERRBB model, implemented in the spectral fitting program XSPEC for stellar BHs (\citealt{ArnaudXPS}).The model describes the observed AD emission computed using a ray-tracing technique in a full relativistic regime, including all the effects such as frame-dragging, Doppler boost, gravitational redshift, light bending, and self-irradiation of the disk (i.e., returning radiation). For all the details of the computations, see the reference work by \citet{Lietal}. \citet{Campiti} explained how to extend these computations to SMBHs using the spectrum peak scaling relations. In this work, we adopt the same fixed parameters, namely hardening factor $f_{\rm col} = 1$, the inclusion of the returning radiation, no inner torque, and no limb-darkening effect.

The key motivation for our work stems from the fact that, for larger spin values, more radiation is emitted close to the equatorial plane (mostly due to light bending) and less is emitted along the disk normal with respect to the SS case (see e.g., \citealt{Campiti}, \citealt{Ishi}): for this reason, the dusty torus could absorb a higher fraction of the total disk luminosity. The distribution of the ratios between the IR (torus) and the optical - UV (AD) observed luminosities could provide statistical constraints on the average torus covering factor and possibly on BH spins.

The structure of the paper is the following: in Sect. \ref{notation} we introduce the notation and describe how the AD radiation angular pattern is modified by relativistic effects using the results of KERRBB (and compare them with those from the SS model). In Sect. \ref{sec:2} we describe the criteria adopted to select an AGN sample suitable for the estimate of IR and optical-UV luminosities. In Sections \ref{DISK} and \ref{TORO}, we present the assumptions made in modeling of the AD and torus emission, the "fitting" procedures adopted to infer the two luminosities from the SEDs, and a discussion about the possible sources of contamination and their uncertainties. Results are discussed in Sect. \ref{results}. A summary and conclusions are presented in Sect. \ref{sec-concl}.

We adopt a flat $\Lambda$CDM cosmology with parameters $H_0=67$ km s$^{-1}$ Mpc$^{-1}$ and $\Omega_{\rm M}=0.32$ (Planck Collaboration, 2018).

%%%%%%%%%%%%%%%%%%%%%%%%%%%%%%%%%%%%%%%%%%%%%%%%%%%%%%%%%%%%%%%%%%%%%%%%%%%
%%%%%%%%%%%%%%%%%%%%%%%%%%%%%%%%%%%%%%%%%%%%%%%%%%%%%%%%%%%%%%%%%%%%%%%%%%%
%%%%%%%%%%%%%%%%%%%%%%%%%%%%%%%%%%%%%%%%%%%%%%%%%%%%%%%%%%%%%%%%%%%%%%%%%%%
%%%%%%%%%%%%%%%%%%%%%%%%%%%%%%%%%%%%%%%%%%%%%%%%%%%%%%%%%%%%%%%%%%%%%%%%%%%

\section{Notation and disk radiation pattern}\label{notation}

The isotropic-equivalent disk luminosity or \textit{observed disk luminosity} $L^{\rm iso}_{\rm d}$ is a quantity derived from the observed flux integrated over the frequency range in which an AD emits its radiation, often identified with the so-called Big Blue Bump (BBB) in the optical-UV bands (see e.g. \citealt{Cunnin}; \citealt{Caldero}; \citealt{Campiti}) under the over-simplified assumption that the AD emits isotropically, i.e., $L^{\rm iso}_{\rm d} = 4 \pi d^2_{\rm L} \int F_\nu d\nu$ (where $d_{\rm L}$ is the luminosity distance and $F_{\nu}$ is the flux density). This quantity does not correspond to the bolometric luminosity $L_{\rm bol}$, used in spectroscopic studies (e.g., \citealt{Shenetal11}) to estimate the accretion power output from the monochromatic luminosity at a specific wavelength, using a bolometric correction (e.g., \citealt{Rich06}). In general, $L_{\rm bol}$ normally also includes the IR and the X-ray emissions produced by the dusty torus and the X-ray corona (i.e., AD reprocessed and up-scattered radiation). \citet{Caldero} derived that on average $L_{\rm bol} \sim 2 L^{\rm iso}_{\rm d}$. 

\begin{figure}
\centering
\hskip -0.2 cm
\includegraphics[width=0.5\textwidth]{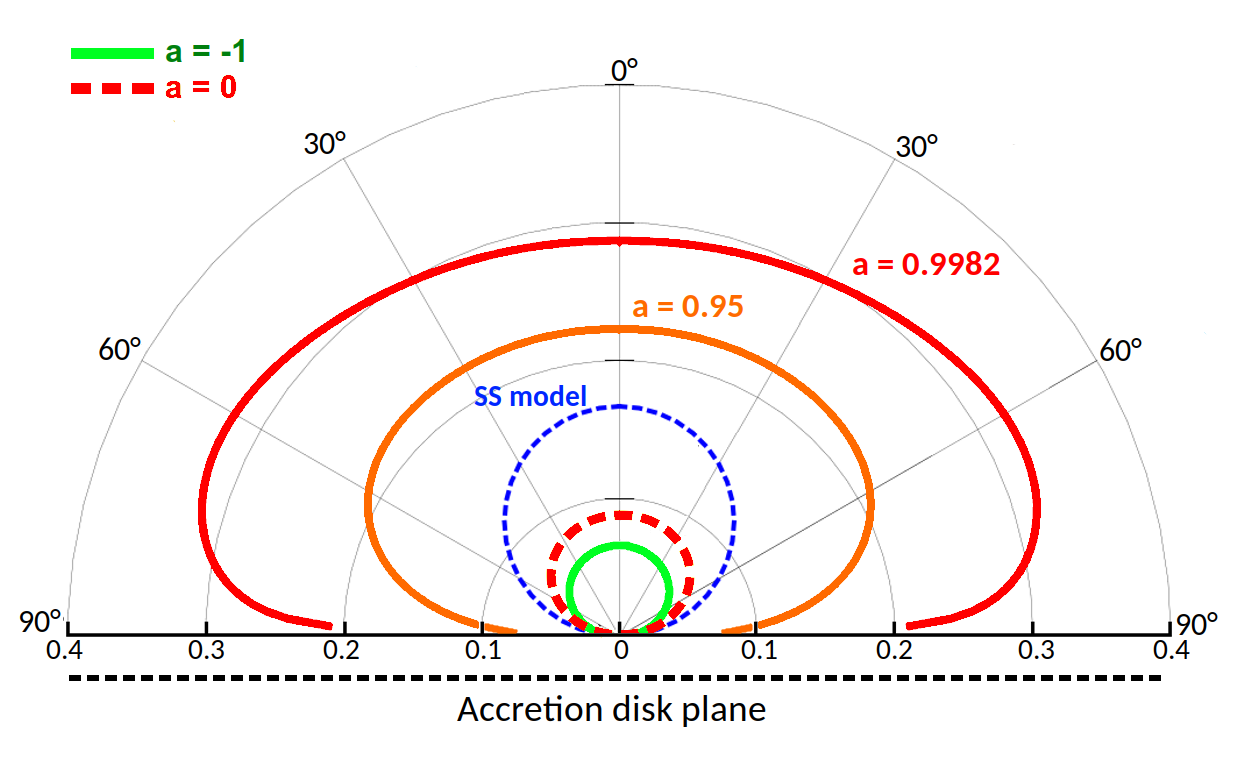}
\caption{Relativistic AD radiation angular pattern in polar coordinates for different spin values ($a=-1$ green line, $a=0$ light brown line, $a=0.95$ orange line, $a=0.9982$ red line) calculated with KERRBB and compared with the non-relativistic SS model (dashed blue line). The radial axis is the isotropic disk luminosity $L^{\rm iso}_{\rm d}$ normalized to $\dot{M} c^2$ (see text).} 
\label{gridpattern}
\end{figure}

For the purpose of this work, we need to define some fundamental quantities to be linked to the observational ones. The \textit{total disk luminosity} $L_{\rm d}(a)$ is the total luminosity emitted from the AD and depends on the spin $a$ through the radiative efficiency $\eta$, that is, $L_{\rm d}(a) = \eta(a) \dot{M} c^2$ (where $\dot{M}$ is the accretion rate). For both the SS and KERRBB models, the AD is not emitting isotropically and the observed flux (and hence $L^{\rm iso}_{\rm d}$) strongly depends on the viewing angle $\theta_{\rm v}$ (measured from the disk normal): this dependence can be described by a general function $f(\theta_{\rm v},a)$ such that $L^{\rm iso}_{\rm d} = f(\theta_{\rm v},a) L_{\rm d}(a)$ (with the normalization $\int^{\pi/2}_{0} f(\theta_{\rm v}, a) \sin \theta\ d \theta = 1$). For a disk described by the SS model, we have $f(\theta_{\rm v}) = 2 \cos \theta_{\rm v}$ (corresponding to the Newtonian case; see e.g., \citealt{Cunnin}; \citealt{Caldero}) while for KERRBB, \citet{Campiti} found an analytical function for $f(\theta_{\rm v},a)$: for large spin values, relativistic effects (mostly light bending) lead to larger AD luminosities at larger viewing angles (e.g., \citealt{Campiti}; \citealt{Ishi}), contrary to the $\cos \theta_{\rm v}$ pattern followed by the SS model. Figure \ref{gridpattern} shows the emission pattern for both models and different spin values: we note that the SS disk case (dashed blue line) is not equivalent to the KERRBB $a = 0$ case because there are relativistic effects neglected in the SS simplified treatment; it is, however, instructive to consider the SS case for comparison with other studies in the literature. 

In the $\nu - \nu L_{\nu}$ representation, $L^{\rm iso}_{\rm d}$ can be inferred directly from the spectral UV peak luminosity $\nu_{\rm p} L_{\nu_{\rm p}}$ which can be constrained with a disk model: for both the SS model (see \citealt{Caldero}) and the relativistic KERRBB model (see \citealt{Campiti}), the spectral shape is almost invariant for different combinations of the parameters (i.e., $M$, $\dot{M}$, $a$, $\theta_{\rm v}$) and a good approximation is $L^{\rm iso}_{\rm d} \sim 2 \nu_{\rm p} L_{\nu_{\rm p}}$ (with an accuracy of $\sim 5 \%$ for all spin values).\footnote{It is important to note that the radiation pattern function depends on $\nu$ because the emission at different frequencies is produced by different regions of the AD where relativistic effects are different: the spectrum at smaller frequencies (i.e., near-infrared--Optical bands), produced by the outer disk annuli where relativistic effects are negligible, follows the $\sim \cos \theta_{\rm v}$ pattern; instead, the large frequency emission produced by the inner annuli close to the BH where relativistic effects are stronger follows the same radiation pattern as $L^{\rm iso}_{\rm d}$ (as this latter is proportional to $\nu_{\rm p}L_{\nu_{\rm p}}$) and strongly depends on the BH spin as shown in the right panel of Fig. \ref{gridpattern} (see \citealt{Campiti}, Fig. 3).}

\begin{figure*}
\centering
\hskip -0.2 cm
\includegraphics[width=0.49\textwidth]{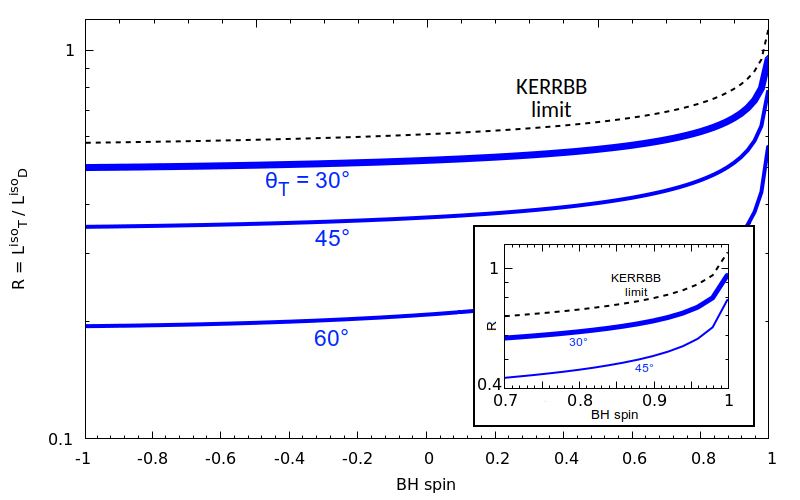} \includegraphics[width=0.49\textwidth]{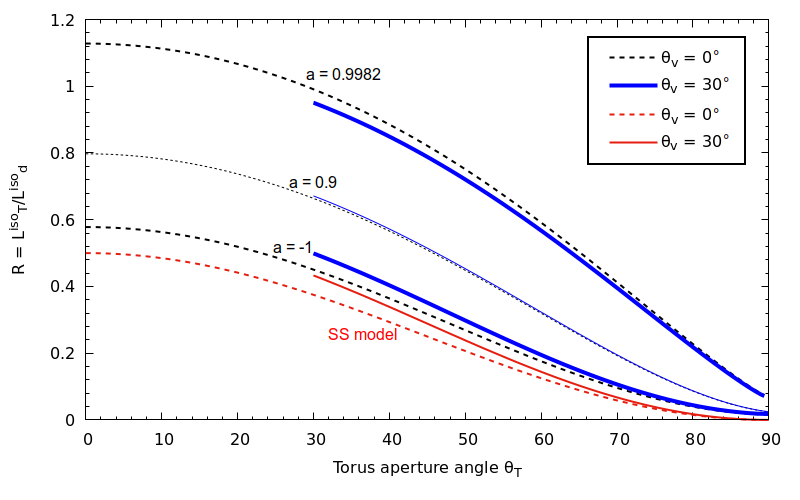}
\caption{Left panel: Luminosity ratio $R$ as a function of the BH spin for different torus aperture angles ($\theta_{\rm T} = 30^{\circ} - 45^{\circ} - 60^{\circ}$) and for $\theta_{\rm v} = 30^{\circ}$ (no solution is shown for $\theta_{\rm T} < \theta_{\rm v}$). The dashed black line is the KERRBB limit ($\theta_{\rm v} = \theta_{\rm T} = 0^{\circ}$). In the SS case, the ratios (not shown for clarity) are $R = 0.43 - 0.29 - 0.14$ for $\theta_{\rm T} = 30^{\circ} - 45^{\circ} - 60^{\circ}$, respectively (see right panel). The small plot is a zoom onto the $a>0.7$ region. Right panel: Luminosity ratio $R$ as a function of the torus aperture angle $\theta_{\rm T}$, for different spin values ($a=-1, 0.9, 0.9982$) and viewing angles ($\theta_{\rm v} = 0^{\circ} - 30^{\circ}$). The $a=0$ case is similar to the one with $a=-1$. All solutions lie between the two extreme spin curves ($a=-1, 0.9982$). For comparison, the red curves represent the SS results. In both plots, the curves are calculated using Eq. \ref{eq_ratio}.} 
\label{gridpattern2}
\end{figure*}

The torus intercepts part of the disk radiation depending on its aperture angle $\theta_{\rm T}$ (measured from the disk normal): the \textit{total torus luminosity} $L_{\rm T}$ depends on $\theta_{\rm T}$ and the BH spin $a$. For a toroidal structure, $\theta_{\rm T} \geq \theta_{\rm v}$ in order to see the AD. Therefore, this quantity can be defined as:
\begin{equation}\label{Equatoro}
	L_{\rm T}(\theta_{\rm T}, a) = L_{\rm d}(a)\ \underbrace{\int^{\pi/2}_{\theta_{\rm T}} f(\theta, a) \sin \theta d \theta}_{= \mathcal{I}(\theta_{\rm T},a)}
\end{equation}

The integral $\mathcal{I}(\theta_{\rm T},a)$ represents the fraction of the total disk luminosity absorbed and re-processed by the torus: for the SS model $\mathcal{I}(\theta_{\rm T}) = \cos^2 \theta_{\rm T}$ because there is no dependence on $a$, while for KERRBB the integral must be solved numerically given its dependence on the BH spin. For simplicity, the torus is assumed to emit isotropically thus, the isotropic equivalent torus luminosity or \textit{observed torus luminosity} $L^{\rm iso}_{\rm T}$ is equivalent to Eq. \ref{Equatoro} (for a discussion, see Sect. \ref{TORO}).

The crucial quantity that we investigate here is the \textit{luminosity ratio} $R$, defined as the ratio between the isotropic equivalent luminosities (i.e., estimated using the observed luminosities):
\begin{equation}\label{eq_ratio}
	R = \frac{L^{\rm iso}_{\rm T}}{L^{\rm iso}_{\rm d}} = \frac{L_{\rm d} \mathcal{I}(\theta_{\rm T}, a)}{L_{\rm d}\ f(\theta_{\rm v}, a)} = \begin{dcases} \frac{\cos^2 \theta_{\rm T}}{2 \cos \theta_{\rm v}}\ \hspace{3mm} ({\rm SS}) \\ \\ \frac{\mathcal{I}(\theta_{\rm T}, a)}{f(\theta_{\rm v}, a)}\ \hspace{3mm} ({\rm KERRBB}) \end{dcases}
\end{equation}

\begin{figure}
\centering
\hskip -0.2 cm
\includegraphics[width=0.49\textwidth]{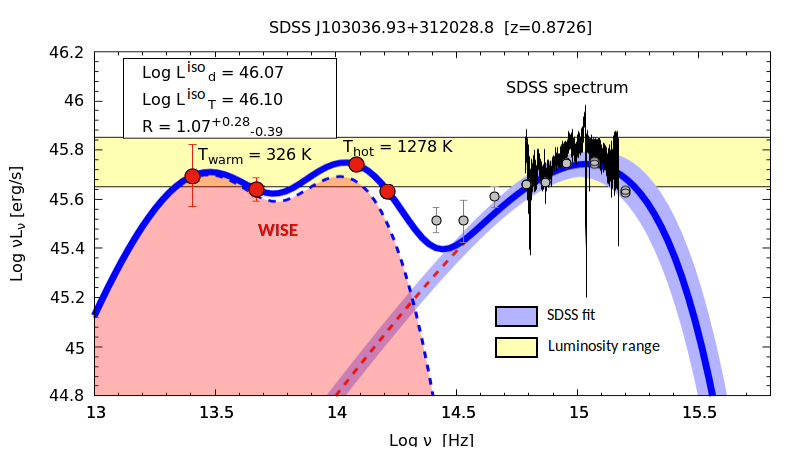}
\caption{Example of SED modeling. The SDSS spectrum (black line) continuum is described with the KERRBB model (dashed red line) while the torus emission is constrained using the four WISE data points (red dots) and two black bodies (dashed blue line contour) plotted along with the corresponding temperatures. The thick blue line is the overall model (disk + torus). We report the isotropic disk and torus luminosities (in erg/s) and the luminosity ratio $R$. Some archival photometric data (2MASS, NED, GALEX - gray dots) are added to the plot. The yellow shaded area is the luminosity range in which $\nu_{\rm p} L_{\nu_{\rm p}}$ lies, which is obtained whilst taking into account different uncertainties. For details about the fitting procedure, the uncertainties, and the constraints on $\theta_{\rm T}$ and $a$, see Sects. \ref{DISK} and \ref{TORO}.} 
\label{plot_example_1}
\end{figure}

For the SS case, for $\theta_{\rm v} = 0^{\circ}$, we have $L^{\rm iso}_{\rm d} = 2 L_{\rm d}$ and therefore the maximum value for $R$ is $0.5$, corresponding to a torus covering the disk completely (i.e., $\theta_{\rm T} = 0^{\circ}$). Instead, the behavior in the KERRBB case is strikingly different: by increasing the spin, more radiation close to the equatorial plane and less is emitted along the disk normal; this makes the torus intercept a larger fraction of $L_{\rm d}$ (Fig. \ref{gridpattern}) resulting in ratios even larger than 1 (Fig. \ref{gridpattern2}). As an example, Fig. \ref{plot_example_1} shows the IR-to-UV SED modeling of one of the sources analyzed in this work (for details about the fitting procedure and uncertainties, see Sect. \ref{DISK} and \ref{TORO}): the observed ratio is $R \sim 1.07$ which, for $\theta_{\rm v} = 0^{\circ}$ and $\theta_{\rm T} = 20^{\circ}$, corresponds to a maximally spinning BH ($a = 0.9982$; Fig. \ref{gridpattern2}, right panel); instead, for $\theta_{\rm v} = 30^{\circ}$, the whole interval of $R$ constrains the BH spin to $a>0.9$ and the torus aperture angle $\theta_{\rm T} < 55^{\circ}$. Therefore, in general, the constraints on the BH spin become tighter for large luminosity ratios (Fig. \ref{gridpattern2}, left panel), while for small values, constraints can be set only for the torus aperture angle (Fig. \ref{gridpattern2}, right panel).

\subsection{Close-to-Eddington accretion} \label{Eddedd}

The strength of the disk radiation is often associated with the Eddington ratio, defined as $\lambda_{\rm Edd} = L_{\rm d} / L_{\rm Edd}$, where $L_{\rm Edd}$ is the Eddington luminosity: for the SS model, given a fixed viewing angle, $M$ and $\dot{M}$ (and thus $L_{\rm d}$ and $\lambda_{\rm Edd}$) are uniquely found by knowing the spectral peak frequency $\nu_{\rm p}$ and luminosity $\nu_{\rm p} L_{\nu_{\rm p}}$ (in a $\nu - \nu L_{\nu}$ representation; see e.g., \citealt{Caldero}); for KERRBB, the dependence of the spectrum peak position on the BH spin induces a degeneracy in the estimates of $M$ and $\dot{M}$ and thus also in the estimates of the total disk luminosity and the Eddington ratio (see e.g., \citealt{Campiti, Campitib}). Regarding $\lambda_{\rm Edd}$, it is important to mark the fact that for large values, the thin disk approximation implemented in KERRBB is not physically correct: for $\lambda_{\rm Edd} \geq 0.3$ (see e.g., \citealt{LaoNet}; \citealt{Kora}; \citealt{McClint}), the disk inflates due to the radiation pressure, and other models must be used (the so-called `slim' or `thick' regime). To study whether or not the nature of the disk influences its radiation pattern, we considered the case of a slim disk and used the relativistic AD model SLIMBH (e.g., \citealt{Sad09}; \citealt{SadwAbra09,SadwAbra}), implemented in XSPEC, to compute $R$. Similarly to what was done for KERRBB, it is possible to find an analytical approximation for the emission pattern (see also \citealt{Campitib}): the theoretical SLIMBH values of $R$ differ from those found with KERRBB by a factor of $< 5\%$ for all angles, spins and Eddington ratios (in the range $0.01 < \lambda_{\rm Edd} < 1$), and for this reason we use the KERRBB results and approximations throughout the paper.

\begin{figure*}
\centering
\hskip -0.2 cm
\includegraphics[width=0.49\textwidth]{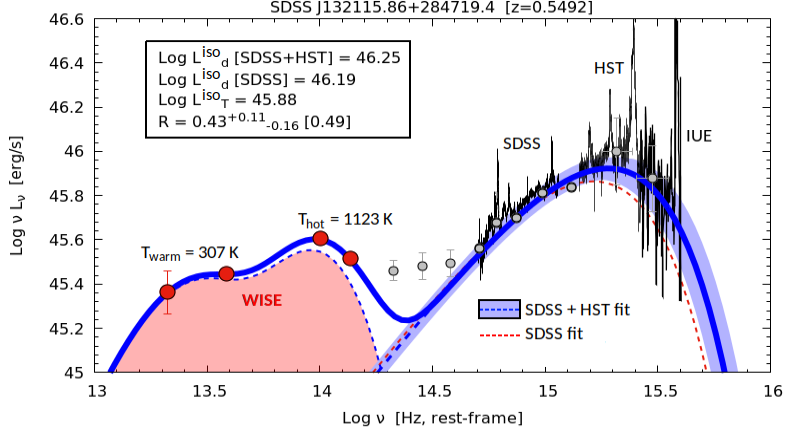}\includegraphics[width=0.48\textwidth]{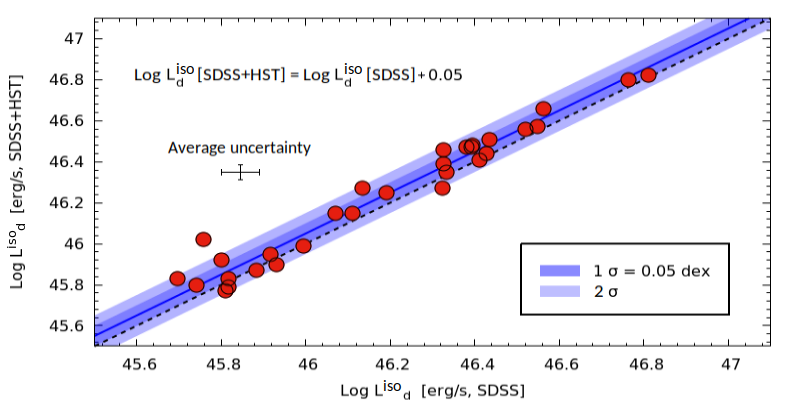} 
\caption{Left panel: Example of the fit of the composite SDSS+HST+IUE spectrum (black lines) for one of the sources of the SDSS+HST sample. The disk emission is modeled with KERRBB (dashed blue line), compared with the fit performed only with the SDSS spectrum (dashed red line). The shaded blue area is a confidence interval of $\sim 0.05$ dex on the spectrum peak. The luminosities inferred from both the fits are reported on the plot (in erg/s) along with the luminosity ratio $R$ (the value computed by considering only the SDSS spectrum is reported n brackets). The fit of the four WISE data points (red dots) is performed with two black bodies (shaded red area with a dashed line contour) plotted along with the corresponding temperatures. The thick blue line is the overall model (disk + torus). Some archival photometric data (2MASS, NED, GALEX - gray dots) are added to the plot (not used in the fitting process). Right panel: Comparison between the isotropic disk luminosities computed from the fit of the SDSS spectrum alone and those computed using additional HST data. The average uncertainty from the fit is shown in the plot. The best-fit relation (blue line) is reported with the 1-2 $\sigma$ data dispersion (shown with shaded blue areas) and the 1:1 line (dashed black line).}
\label{LBOL_new_100}
\end{figure*}

%%%%%%%%%%%%%%%%%%%%%%%%%%%%%%%%%%%%%%%%%%%%%%%%%%
%%%%%%%%%%%%%%%%%%%%%%%%%%%%%%%%%%%%%%%%%%%%%%%%%%
%%%%%%%%%%%%%%%%%%%%%%%%%%%%%%%%%%%%%%%%%%%%%%%%%%
%%%%%%%%%%%%%%%%%%%%%%%%%%%%%%%%%%%%%%%%%%%%%%%%%%
%%%%%%%%%%%%%%%%%%%%%%%%%%%%%%%%%%%%%%%%%%%%%%%%%%
%%%%%%%%%%%%%%%%%%%%%%%%%%%%%%%%%%%%%%%%%%%%%%%%%%
%%%%%%%%%%%%%%%%%%%%%%%%%%%%%%%%%%%%%%%%%%%%%%%%%%

\section{The sample} \label{sec:2}

We considered the SDSS DR7Q catalog (containing 105,783 QSOs) whose continuum and line luminosities have already been estimated and studied by \citet{Shenetal11}. In this catalog, all the most common AGN emission lines have a full width at half maximum FWHM $>1000$ km/s, and therefore all QSOs can be classified as Type 1 (e.g., \citealt{Anto}). All spectra are obtained in the observed wavelength range $3800-9200$ \AA. To define a suitable sample, we adopted the following selection criteria:
\begin{itemize}
	\item We required that the sources have a measured rest-frame monochromatic luminosity, both at 3000\AA\ and 5100\AA\ , to estimate the continuum slope. As the AD emission is identified with the BBB, we excluded sources with no evidence of such a feature by requiring that the spectral slope be positive (in the $\nu - \nu L_{\nu}$ representation). Given the limited wavelength coverage of the SDSS spectrum, the selected sources have a redshift in the range $0.35 < z < 0.89$. This criterium set a lower limit for the spectrum peak frequency, Log $\nu_{\rm p} / {\rm Hz} \geq 14.9$: sources hosting very massive BHs are possibly neglected.\footnote{The average BH mass computed with KERRBB for the whole sample is Log $M / M_{\odot} = 9.00 \pm 0.20$ and the lower limit for the peak frequency (Log $\nu_{\rm p}/ {\rm Hz} > 14.9$) led to sources with masses of Log $M / M_{\odot} \gsim 9.5$ being neglected.}
	\item A further criterium was imposed to minimize the host galaxy contamination in the optical band. Following \citet{Shenetal11}, we selected sources with a SDSS bolometric luminosity $L_{\rm bol}$ $\gsim 10^{46}$ erg/s: for those sources, the SDSS monochromatic luminosity at 5100\AA\ is Log $L_{5100} \gsim 45$ erg/s, and is contaminated by the galactic emission by a factor of $\sim 5 \%$ and smaller at lower wavelengths. In this way, we reduced the number of components in the fitting procedure by neglecting the contribution from the host galaxy.
	\item To estimate the dust-reprocessed emission in the IR band, we cross-correlated the previously selected sources with the Wide-field Infrared Survey Explorer (WISE; \citealt{Wright}) catalog, selecting only those with detection in all of the four WISE IR bands (3.4, 4.6, 12 and 22 $\mu$m). 
	\item Finally, to avoid possible contamination of the IR and UV bands from synchrotron emission, the cross-correlation between the SDSS DR7Q catalog and the Faint Images of the Radio Sky at Twenty-centimeter survey (FIRST; \citealt{Becker}) reported by \citet{Shenetal11} allowed us to select only radio-quiet sources: following the definition of radio-loudness $\mathcal{R L}$ adopted by \citet{Shenetal11},\footnote{The definition of radio loudness is defined as $\mathcal{R L} = F_{\nu, {\rm 6\ cm}} / F_{\nu, 2500{\rm \AA}}$, where $F_{\nu, {\rm b}}$ is the flux density at the wavelength $b$ (\citealt{Shenetal11}).} we selected only the sources with $\mathcal{R L} < 10$ (e.g., \citealt{Keller}), including sources observed by FIRST but without a detectable radio flux, all classified as radio-quiet.\footnote{The flux limit of FIRST is $\sim$1 mJy at 1.4 GHz (\citealt{Becker}).}
\end{itemize}

The resulting final sample comprises 2922 sources. A further selection criterium is applied in the following section to consider only sources with a good fit of the SDSS spectrum.

%%%%%%%%%%%%%%%%%%%%%%%%%%%%%%%%%%%%%%%%%%%%%%%%%%%%%%%%%%%%%%%%%%%%%%%%%%%
%%%%%%%%%%%%%%%%%%%%%%%%%%%%%%%%%%%%%%%%%%%%%%%%%%%%%%%%%%%%%%%%%%%%%%%%%%%
%%%%%%%%%%%%%%%%%%%%%%%%%%%%%%%%%%%%%%%%%%%%%%%%%%%%%%%%%%%%%%%%%%%%%%%%%%%
%%%%%%%%%%%%%%%%%%%%%%%%%%%%%%%%%%%%%%%%%%%%%%%%%%%%%%%%%%%%%%%%%%%%%%%%%%%

\section{Accretion disk}\label{DISK}

In this section we describe the fitting procedures adopted to infer $L^{\rm iso}_{\rm d}$ and discuss the possible sources of uncertainty. It is important to stress that the fitting procedure is simply designed to determine $L^{\rm iso}_{\rm d}$ (constrained from the SED) and not to constrain the physical parameters of the sources.

%%%%%%%%%%%%%%%%%%%%%%%%%%%%%%%%%%%%%%%%%%%%%%%%%%%%%%%%%%%%%%%%%%%%%%%%%%%%%%%%

\subsection{Emission}\label{ADemission}

The AD emission is identified with the BBB component in the optical-UV band. Given that all sources are Type 1 QSOs, we assume that the isotropic disk luminosity $L^{\rm iso}_{\rm d}$ is free from absorption by the torus. We modeled the AGN continuum of the AD by fitting the SDSS spectrum with KERRBB: using the fact that the KERRBB spectral shape is almost invariant for different parameter combinations (\citealt{Campiti}), we performed the fit using GNUPLOT (which includes a non-linear least-squares Marquardt-Levenberg algorithm), using the curvature of the SDSS spectrum to constrain the peak frequency $\nu_{\rm p}$ and luminosity $\nu_{\rm p}L_{\nu_{\rm p}}$ (and thus $L^{\rm iso}_{\rm d}$) even if the peak is not covered by the SDSS data. In the fitting procedure, we did not include (1) emission or absorption lines, because those spectral features have no drastic effects on the overall KERRBB fit, or (2) possible available photometric data, because they could be contaminated by some emission or absorption lines.

The SDSS spectrum shows a superimposed minor component named "Small Blue Bump" (from Log $\nu / {\rm Hz} = 14.9 - 15.1$, rest-frame), caused by the blending of several iron lines and the hydrogen Balmer continuum (\citealt{Wills}; \citealt{Vanden}), which does not affect the localization of the spectrum peak: even though the SDSS spectral coverage is limited, the curvature of the available spectrum can still be used to constrain the peak (in the frequency range Log $\nu_{\rm p} / {\rm Hz} \sim 15 - 15.5$, rest-frame; e.g. \citealt{Campitic}). 

For the analyses performed in the following sections, we chose only the sources with the best peak position estimation: we selected only those sources with an uncertainty on both $\nu_{\rm p}$ and $\nu_{\rm p}L_{\nu_{\rm p}}$ less than $\sim 0.05$ dex. This criterium reduces our initial sample to 1858 sources (hereafter "SDSS sample").

As a further check of the degree to which the spectral curvature at lower frequencies provides significant constraints on $\nu_{\rm p}$ and $\nu_{\rm p}L_{\nu_{\rm p}}$ (and so on $L^{\rm iso}_{\rm d}$), we cross-matched our sample with the HST catalog and built a subsample of 30 objects (hereafter, "SDSS+HST sample") with UV spectroscopic data for a wider wavelength range (Log $\nu / {\rm Hz} = 14.7 - 15.6$, rest-frame); we also included in the fits available data from the Far Ultraviolet Spectroscopic Explorer (FUSE) for five sources and data from the International Ultraviolet Explorer (IUE) for two sources (such as the one in Fig. \ref{LBOL_new_100}).\footnote{All spectroscopic data were retrieved from the online Mikulski Archive for Space Telescopes (MAST).} We corrected the data from the Galactic extinction using the \citet{Cardelli} reddening law and $E_{\rm B-V}$ from the map of \citet{Schle} with an extinction factor $R_{\rm V} = 3.1$. When possible, the non-simultaneous spectra were calibrated by matching the flux in their common wavelength ranges, assuming that the spectral shape does not change during flux variations (the maximum flux mismatch we found is less than $\sim 20 \%$; see e.g., \citealt{Shang05}). 

\begin{figure*}
\centering
\hskip -0.2 cm
\includegraphics[width=0.51\textwidth]{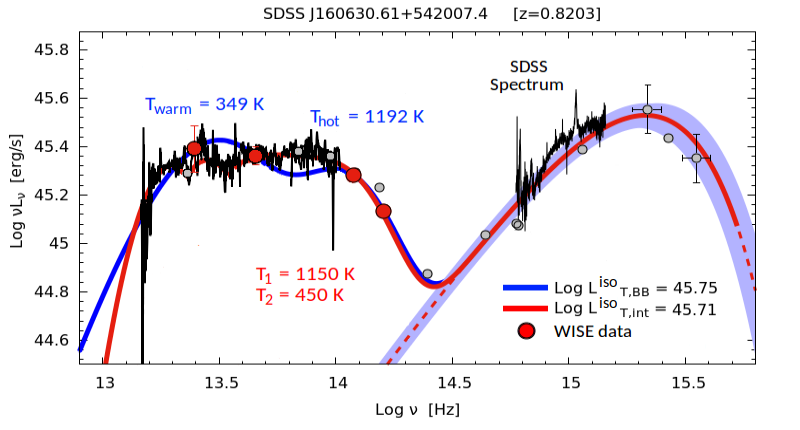}\includegraphics[width=0.47\textwidth]{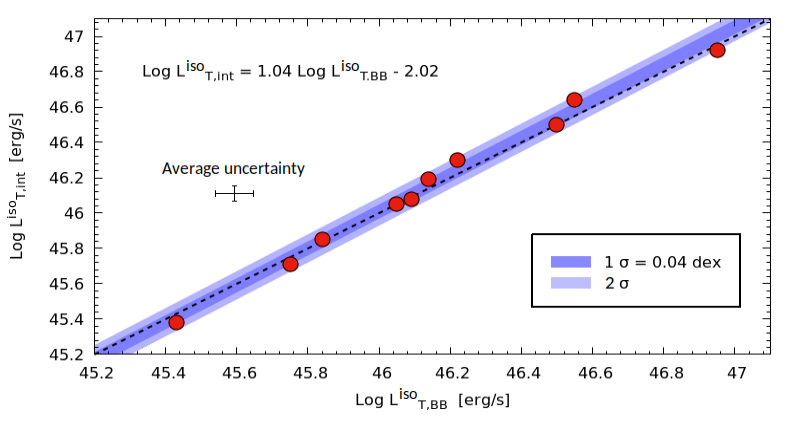}
\caption{Left panel: Example of fit of the SDSS optical - UV and SPITZER IR spectra (black lines). The thick blue line is the fit performed with two black-bodies (whose temperatures, $T_{\rm warm}$ and $T_{\rm hot}$, are reported in the plot) using the four WISE data points (red points) for the IR emission, and KERRBB for the disk emission (dashed red line with a shaded blue area representing the confidence interval for the spectrum peak - $\sim 0.05$ dex). The thick red line is the fit performed with two parabolas and two black-bodies (whose temperatures, $T_{1}$ and $T_2$, are reported on the plot; see text for details) to describe the SPITZER emission, and KERRBB for the disk emission. Both the integrated luminosities ($L^{\rm iso}_{\rm T, BB}$ and $L^{\rm iso}_{\rm T, int}$) are reported on the plot (in erg/s). Archival photometric data (2MASS, NED, GALEX) are shown with gray dots. Right panel: Comparison between $L^{\rm iso}_{\rm T,BB}$ and $L^{\rm iso}_{\rm T,int}$ for the sources with SPITZER IR data. The average uncertainty from the fit is shown in the plot. The best-fit equation is reported along with the 1-2$\sigma$ data dispersion (shaded blue areas) and the 1:1 line (dashed black line).}
\label{plot_example_SPITZ}
\end{figure*}

From the fit of the SDSS+HST sample sources, we found that $\nu_{\rm p}L_{\nu_{\rm p}}$ is on average higher by a factor of $\sim 0.05$ dex with respect to the one obtained from the fit of the SDSS spectrum alone (Fig. \ref{LBOL_new_100}). As we used spectroscopic data in the FUV band, we checked the possibility that our results might be affected by the presence of blended interstellar absorption features (at Log $\nu / {\rm Hz} > 15.4$) which could reduce the AGN continuum flux. To quantify this effect, we performed the same fitting procedure using only spectroscopic data at Log $\nu / {\rm Hz} < 15.4$ (assuming that this frequency range is free from absorption) and found that the values of $\nu_{\rm p} L_{\nu_{\rm p}}$ are consistent with the previous estimates within an interval of $< 0.05$ dex. We consider these as typical uncertainties for the SDSS sample.

%%%%%%%%%%%%%%%%%%%%%%%%%%%%%%%%%%%%%%%%%%%%%%%%%%%%%%%%%%%%%%%%%%%%%%%%%%%%%%%%

\subsection{Caveats} \label{cavo}

Some structures close to the AD as well as dust located along the line of sight can lead to incorrect estimates of $\nu_{\rm p}L_{\nu_{\rm p}}$ and $L^{\rm iso}_{\rm d}$. Here we discuss these in order to define a confidence interval for the ratio $R$ in the following section. 

\textit{Dust and intrinsic absorption}: To estimate the effects of dust absorption, we followed the procedure detailed in \citet{Campitic}. For the redshift range spanned by our sample, we estimate that the UV attenuation due to the intergalactic medium is negligible (see \citealt{Madau}; \citealt{Haamad}; see also \citealt{Castietal}). For what concerns the interstellar medium of the host galaxy, following \citet{Baron}, we find that $\sim 80 \%$ of the sources show an extinction of $E_{\rm B-V} \lsim 0.05$ mag (the average value is $\sim 0.03$ mag), which is computed using the SDSS spectral slope and is consistent with what is thought to be the value for Type 1 AGNs ($E_{\rm B-V} < 0.1$ mag; e.g., \citealt{Kora}). Consequently, the extinction-corrected AD luminosities would be larger (on average) by a factor of $\lsim 0.1$ dex. Given the uncertainties involved in this procedure and since possible changes in the UV slope could be caused by other factors connected to the BH physics (i.e., mass, accretion rate, spin; see e.g., \citealt{Hub2000}; \citealt{DavLao}), we did not consider any dust correction.

\textit{X-ray corona}: This structure is located above and close to the inner region of the AD. Different geometries have been proposed (e.g., lamp post - \citealt{Miniu}, extended slab or sphere - \citealt{Petru}; \citealt{Chaina}; \citealt{Done}). Assuming that it up-scatters part of the AD radiation in the X band (e.g., \citealt{Sazo}; \citealt{LuRi}), the intrinsic $L^{\rm iso}_{\rm d}$ would be larger with respect to the observed one, resulting in a smaller $R$. Following the work of \citet{Duras} (see also e.g., \citealt{VasuFa}; \citealt{Lusso}), sources with a bolometric luminosity $L_{\rm bol} > 10^{46}$ erg/s have a X-band luminosity of $L_{\rm X} < 0.1\ L_{\rm bol}$: assuming that the bolometric luminosity is $L_{\rm bol} \sim 2\ L^{\rm iso}_{\rm d}$ (as found by \citealt{Caldero}), the X-band luminosity is $L_{\rm X} < 0.2 L^{\rm obs}_{\rm d}$ on average. If this latter fraction ($<0.2$) corresponds to the fraction of disk radiation up-scattered by the corona in the X band, the intrinsic $L^{\rm iso}_{\rm d}$ would be larger by a factor of $< 0.08$ dex, leading to a smaller $R$ by the same amount which can be considered as average upper limits.\footnote{The effect of the X-ray Corona on the UV emission can be studied using sophisticated broad-band models (e.g., OPTXAGNF, \citealt{Done}; AGNSED, \citealt{KuDo}). Unfortunately, for our sample, the majority of the sources has a limited data coverage which cannot allow us to use those models appropriately; moreover, simultaneous data are necessary to perform a proper parameter estimation.}

\textit{Variability}: Disk flux changes could modify both $L^{\rm obs}_{\rm d}$ and $L^{\rm obs}_{\rm T}$ given that this latter is proportional to the disk luminosity. In this context, the time-lag between the AD and the torus is important: for bright sources ($L_{\rm bol} > 10^{46}$ erg/s), the sublimation radius of the toroidal dust is located at a few parsecs from the SMBH (e.g., \citealt{Barva}), and therefore IR flux variations are expected to occur a few years after the disk ones (e.g., \citealt{Lyu19}). For this reason, in variable sources and in short time intervals, only the disk luminosity can be observed to change while the luminosity of the torus remains constant. Assuming a flux variability of $0.1$ dex, the disk luminosity would change by the same amount. However, from a statistical point of view, only a few sources are expected to vary by a significant amount, and in large samples this effect can be negligible. For this reason, disk variability was not considered in this work.

%%%%%%%%%%%%%%%%%%%%%%%%%%%%%%%%%%%%%%%%%%%%%%%%%%%%%%%%%%%%%%%%%%%%%%%%%%%
%%%%%%%%%%%%%%%%%%%%%%%%%%%%%%%%%%%%%%%%%%%%%%%%%%%%%%%%%%%%%%%%%%%%%%%%%%%
%%%%%%%%%%%%%%%%%%%%%%%%%%%%%%%%%%%%%%%%%%%%%%%%%%%%%%%%%%%%%%%%%%%%%%%%%%%
%%%%%%%%%%%%%%%%%%%%%%%%%%%%%%%%%%%%%%%%%%%%%%%%%%%%%%%%%%%%%%%%%%%%%%%%%%%

\section{Torus}\label{TORO}

Here we describe the assumptions and the procedure adopted to estimate the isotropic torus luminosity and quantify the uncertainties involved.

%%%%%%%%%%%%%%%%%%%%%%%%%%%%%%%%%%%%%%%%%%%%%%%%%%%%%%%%%%%%%%%%%%%%%%%%%%%%%%%%

\subsection{Emission} \label{TORUS_struc_emis}

We estimated the torus emission based on the following simple assumptions: (a) the dust is distributed with a symmetric, equatorial structure with an aperture angle $\theta_{\rm T} \geq \theta_{\rm v}$, (b) the disk radiation intercepted by the torus is re-processed and totally re-emitted isotropically, and (c) the torus is assumed to have a continuous dust distribution even though the possible clumpiness (confirmed by observations) could affect the observed torus flux (which results in anisotropic emission as also shown by several numerical models; see e.g., \citealt{Nenka,Nenkb}). A discussion about the effects of an angle-dependent torus emission on our results is presented in Sect. \ref{cavoT}.

The torus emission is estimated from the SED in the frequency range Log $\nu / {\rm Hz} \sim 13 - 14.5$ (rest-frame), as constrained by the four WISE data points. Given that only its luminosity is necessary for our analysis, we simply used two independent black bodies to describe the torus SED instead of sophisticated numerical models (e.g., CLUMPY, \citealt{Nenka, Nenkb}; or CAT3D, \citealt{HonKish}). In this procedure, we assumed an isotropic emission even though several numerical torus models show an angle and frequency-dependent IR emission. Given that those models fail to describe the far-infrared (FIR) emission (peaking at Log $\nu / {\rm Hz} \sim 14$)\footnote{This peak probably originates from the hot dust closer to the SMBH which is described with an additional black body (e.g., \citealt{Deoetal}; \citealt{MorNet}; \citealt{Leip}; \citealt{Krog}; \citealt{Zhuang}).}, in this statistical work, we adopted the simplest model (i.e., isotropic emission and two black bodies).

\begin{figure*}
\centering
\hskip -0.2 cm
\includegraphics[width=0.495\textwidth]{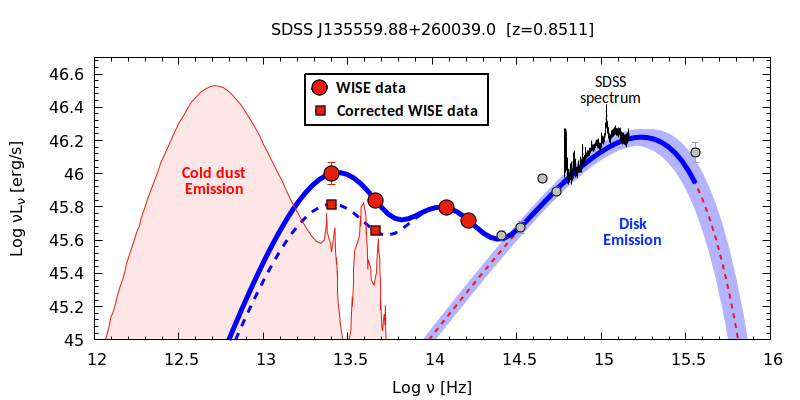}\includegraphics[width=0.495\textwidth]{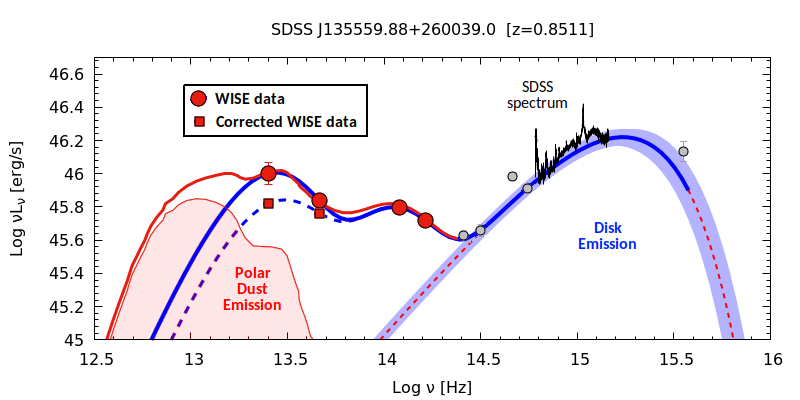} 
\caption{Example of flux correction from cold dust (left panel) and polar dust (right panel) of the WISE data. We used the template of the starburst galaxy M82 (\citealt{Kennetal}, red line) assuming a peak luminosity twice larger than the AD one (described by KERRBB, dashed red line in both panels) and the mean polar dust template as found by \citet{Lyu}, assuming that its contribution to the MIR emission is $\sim 50 \%$ (as found by \citealt{Asmus16} and \citealt{Lyu}). The thick blue line is the disk--torus model related to the uncorrected WISE data and the SDSS spectrum (black line). The dashed blue line is the new fit performed after the correction of the WISE data flux from the contamination. For the cold dust, the uncorrected and corrected luminosity ratios are $R = 0.55$ and $R = 0.44$, respectively; for the polar dust, $R = 0.55$ and $R = 0.39$, respectively. Some archival photometric data (2MASS, NED, GALEX) are plotted with gray dots (not used in the fitting process).}
\label{colddust}
\end{figure*}

The temperature of the two black bodies is set to $T < 2000$ K (i.e., dust sublimation temperature; see e.g., \citealt{Hernan}; \citealt{Caldero12}; \citealt{Collinson}). We used two components to be consistent with the scenario where dust is located at different distances from the SMBH: a hotter component originates from hot (graphite) dust close to the sublimation temperature and facing the disk (e.g., \citealt{Barva}; \citealt{Galla}; \citealt{Mor}) while a colder component originates from the outer region of the torus. This latter emission (characterized by a black-body temperature of $\sim 200 - 400$ K) does not correspond to the cold dust heated by stars (with a temperature $<100$ K; e.g., \citealt{Bendo}; \citealt{Bose}; \citealt{Dale}), and is located at larger distances from the disk and peaking around Log $\nu / {\rm Hz} \sim 12.7$ (e.g., \citealt{Pear}; see also Sect. \ref{cavoT}). Finally, $L^{\rm iso}_{\rm T}$ is obtained as the sum of the luminosities of those two frequency-integrated black bodies.\footnote{We found that the mean temperatures of the hot and warm black bodies are $T_{\rm hot} = 1277$ K and $T_{\rm warm} = 309$ K, respectively, similar to those found by other authors (e.g., \citealt{Hernan}; \citealt{Caldero12}; \citealt{Collinson}). The two black-body luminosities are linked by the relation, $\text{Log}\ L^{\rm iso}_{\rm T, warm} = \text{Log}\ L^{\rm iso}_{\rm T, hot} + 0.11$, with a 1$\sigma$ data dispersion of $\sim 0.17$ dex.}

In order to check the goodness of the two black body approximation, we performed the following analysis: we cross-matched our sample with the SPITZER catalog and found ten sources with IR spectroscopic data in the rest-frame frequency range Log $\nu/ {\rm Hz} = 13 - 14$ and compatible with the WISE photometric data. We described the SPITZER spectrum using two parabolas, one in the rest-frame range Log $\nu/ {\rm Hz} \sim 13 - 13.4$ (describing the cold bump peaking at Log $\nu/ {\rm Hz} \sim 13.3$) and one in the rest-frame range Log $\nu/ {\rm Hz} \sim 13.4 - 13.6$ (corresponding to the silicate emission peaking at Log $\nu/ {\rm Hz} \sim 13.5$ - see e.g., \citealt{HonKish} and references therein), and two black bodies in the range Log $\nu/ {\rm Hz} = 13.6 - 14.5$. We chose parabola-like emission instead of black bodies in order to vary the width of the peak emission and perform a better fit of the SPITZER data in the range Log $\nu/ {\rm Hz} \sim 13 - 13.4$. The colder black body was used to have a better description of the spectrum in the range Log $\nu/ {\rm Hz} = 13.5 - 14$. By integrating all these components in the corresponding frequency ranges (specified above) and summing up their contributions, we obtained the SPITZER torus luminosity $L^{\rm iso}_{\rm T,int}$.

Figure \ref{plot_example_SPITZ} (left panel) shows an example of the fit: the modeling with two black bodies overestimates part of the IR luminosity at Log $\nu/ {\rm Hz} < 13.5$ and underestimates it for Log $\nu/ {\rm Hz} = 13.5 - 14$; on average these two effects balance out resulting in a torus luminosity similar to the one computed by integrating the SPITZER spectrum. Figure \ref{plot_example_SPITZ} (right panel) shows the comparison between the two luminosities: although the sample is rather small, the best fit is consistent with the 1:1 line with a 1$\sigma$ data dispersion of $\sim 0.04$ dex.

%%%%%%%%%%%%%%%%%%%%%%%%%%%%%%%%%%%%%%%%%%%%%%%%%%%%%%%%%%%%%%%%%%%%%%%%%%%%%%%%

\subsection{Infrared contamination and torus anisotropy} \label{cavoT}

Here we discuss the possible sources of contamination that could affect our estimates of $L^{\rm iso}_{\rm T}$ and therefore also the value of the ratio $R$.
	
\textit{Galaxy emission}. We do not expect strong contamination of the estimated torus luminosity from this emission due to the criterium adopted to build our sample (Sect. \ref{sec:2}). However we quantified the effects on $L^{\rm iso}_{\rm T}$ using the galaxy template from \citet{Manuc}: by requiring that $L_{5100}$ is contaminated by a factor of $\sim 5 \%$ at most, we find that the host galaxy SED leads to a modification of $L^{\rm iso}_{\rm T}$ by a negligible factor ($< 2 \%$).

\textit{Cold dust emission}. This emission is related to dust heated by stars with a temperature of $<100$ K (e.g., \citealt{Bendo}; \citealt{Bose}; \citealt{Dale}) and located at larger distances from the AD with respect to the torus. In principle, it could contaminate the cold part of the torus emission leading to an overestimation of $L^{\rm iso}_{\rm T}$. We used the SED of the starburst galaxy M82 (\citealt{Kennetal}) as a template to quantify its effect. Figure \ref{colddust} (left panel) shows the case in which the cold dust peak luminosity is chosen arbitrarily to be twice the luminosity of the AD (as an extreme case). We subtracted its contribution from the WISE data flux and fitted the data with two black bodies: we find that $L^{\rm iso}_{\rm T}$ is overestimated by a factor of $\sim 0.1$ dex leading to a similar modification for the intrinsic $R$. A further test was made using the extinction law found by \citet{Calze} for a sample of local galaxies: the basic assumption is that the galaxy emission in the NIR-Optical bands is attenuated by the host galaxy cold dust; the same amount of absorbed radiation is emitted in the FIR band as a modified black body. Using the law found by \citet{Calze} (their Eqs. 2-3-4) with an intrinsic reddening of $E_{\rm B-V} = 0.1$ mag leads to cold dust emission in the FIR which has a negligible effect on the torus emission ($<2 \%$). It is important to note that these tests are generic analyses due to uncertainties related to modeling of the cold dust emission and the lack of data in the corresponding frequency range for almost the whole sample.

\begin{figure*}
\centering
\hskip -0.2 cm
\includegraphics[width=0.76\textwidth]{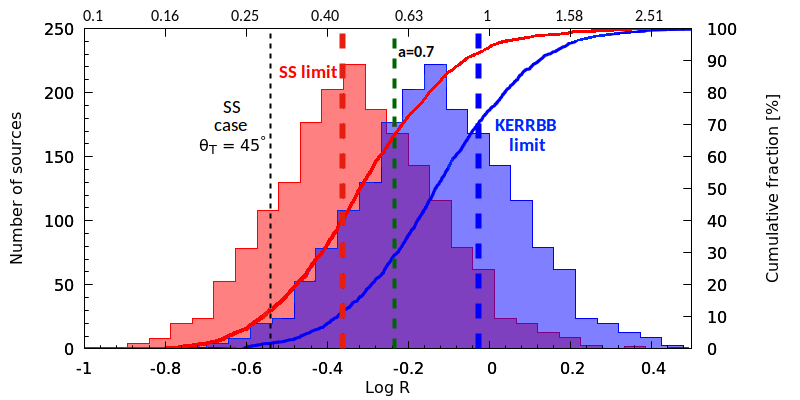}
\caption{Logarithmic distribution of the luminosity ratio $R$ (blue histogram; mean logarithmic value $\langle {\rm Log}\ R \rangle = -0.13 \pm 0.20$). The red histogram is the distribution assuming that $R$ is given by its lower limit (from uncertainties - see Sect. \ref{Runcert}). The solid blue and red lines are the cumulative functions related to the two histograms. The thick dashed blue line is the KERRBB limit ($R = 0.94$, for $\theta_{\rm v} = 30^{\circ}$) while the maximum SS value ($R=0.43$, for $\theta_{\rm v} = 30^{\circ}$) is represented with a dashed red line (the thin dashed black line represents the SS case with $\theta_{\rm T} = 45^{\circ}$ and $\theta_{\rm v} = 30^{\circ}$). Sources whose luminosity ratio $R$ is larger than the one shown with a dashed green line have a BH spin $a>0.7$ (see Sect. \ref{notation}). The top x-axis shows the linear value of $R$.}
\label{ratio_all}
\end{figure*}

\textit{Polar dust}. Recent works (e.g., \citealt{Asmus16}; \citealt{Lopez}; \citealt{Left}; \citealt{Lyu}; \citealt{Asmus19}) have shown that part of the MIR emission originates from polar regions instead of from an equatorial dusty torus with a characteristic dust temperature of $\sim 110$ K (\citealt{Lyu}). In this case, the disk luminosity would be partly absorbed, while the torus would produce less MIR radiation than observed. The results of the works mentioned above suggest a correlation between the disk luminosity, the formation, distance, and extension of polar dust winds,\footnote{Polar dust winds have been observed (e.g., \citealt{Asmus19}) and described by simulations (e.g., \citealt{Vena}) and numerical models (e.g., CAT3D-WIND, \citealt{HonKish17}). For a proper application of such models, more IR data are required for a correct parameter estimation of the main torus features.} and the fraction of MIR emission due to polar dust, although such relations require further investigation. Unfortunately, it was not possible to draw any conclusion regarding those possible relations for any of the sources of our sample. However, using the results of \citet{Asmus19} (their Table 4), it is possible to find that, for the luminosity range spanned by our sample, the contribution of polar dust in the MIR is approximately $\sim 60\% - 70 \%$ (assuming that the relation found in his work holds for all AGNs). In this study, we conducted a general analysis of the possible torus contamination assuming that half of the MIR emission (Log $\nu / {\rm Hz} \sim 13 - 13.5$) from all sources is due to the possible presence of a polar dusty wind (as also found on average by \citealt{Asmus16}):\footnote{The results do not deviate drastically from those reported if a larger fraction ($\sim 60\% - 70 \%$) is used in the calculations.} on average, in such a case, the intrinsic torus luminosity would be dimmer by a factor of $\sim 0.10$ dex with respect to the observed luminosity; moreover, if the polar dust luminosity comes from a reprocessed fraction of the disk luminosity, this latter would be brighter by a factor of $\sim 0.05$ dex with respect to the observed luminosity, leading to an overall modification of the ratio $R$ by a factor of $\sim 0.15$ dex at most (Fig. \ref{colddust}, right panel).\\

\textit{Torus anisotropy}. As already mentioned before, some numerical models show the dependence of the torus luminosity from $\theta_{\rm v}$ (e.g., \citealt{Nenka}; \citealt{HonKish}; \citealt{Stale}), depending also on the dust distribution and its clumpiness, even through they cannot properly fit the IR bump peaking at Log $\nu / {\rm Hz} \sim 14$. \citet{CastiDe} quantify the torus anisotropy though using an analytical expression and the numerical model CAT3D (\citealt{HonKish}): their Eq. 1 represents the angle-dependent observed flux density from which it is possible to find that the observed luminosity is larger by a factor of $a + b \cos \theta_{\rm v}$ (where $a=0.56$ and $b=0.88$) with respect to the intrinsic luminosity. Assuming that the intrinsic torus luminosity is equal to $L_{\rm T}$ (i.e., the fraction of disk radiation absorbed by the torus; Eq. \ref{Equatoro}), the luminosity ratio (Eq. \ref{eq_ratio}) becomes:
\begin{equation}
	R = \frac{L^{\rm iso}_{\rm T}}{L^{\rm iso}_{\rm d}} = \frac{L_{\rm T}\ (a + b \cos \theta_{\rm v})}{L_{\rm d} f(\theta_{\rm v}, a)} = \frac{\mathcal{I}(\theta_{\rm T}, a)\ (a + b \cos \theta_{\rm v})}{f(\theta_{\rm v}, a)}	
\end{equation}

Adopting this correction with $\theta_{\rm v} = 30^{\circ}$, the KERRBB curves plotted in Fig. \ref{gridpattern2} would be shifted towards larger values of $R$ by $\sim 30 \%$. As mentioned above, the approximation found by \citet{CastiDe} depends on the model CAT3D which cannot properly fit the IR bump at Log $\nu / {\rm Hz} \sim 14$, and therefore such a correction has to be taken with care.

%%%%%%%%%%%%%%%%%%%%%%%%%%%%%%%%%%%%%%%%%%%%%%%%%%%%%%%%%%%%%%%%%%%%%%%%%%%%%%%%

\begin{figure*}
\centering
\hskip -0.2 cm
\includegraphics[width=0.505\textwidth]{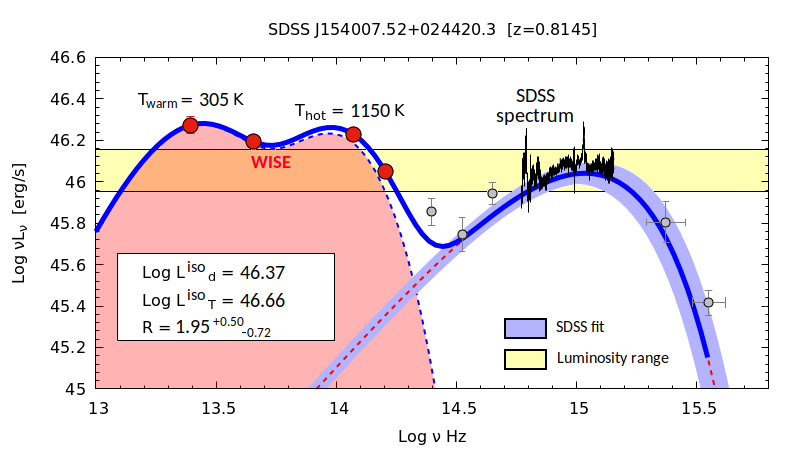}\includegraphics[width=0.505\textwidth]{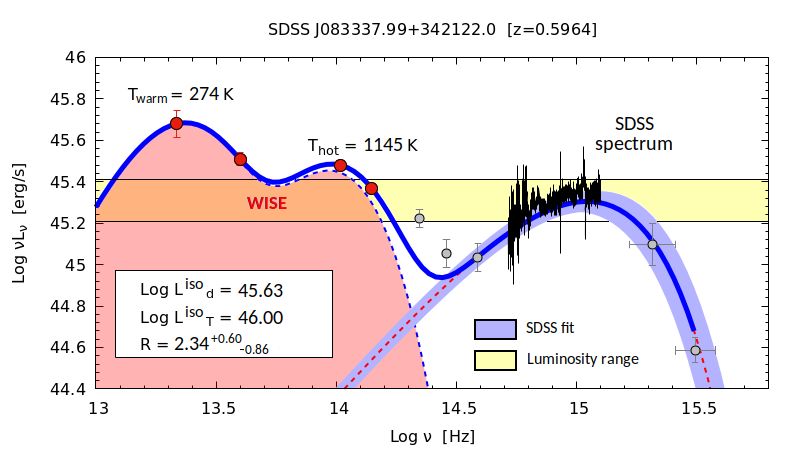} \caption{Fit of two sources with a large luminosity ratio that cannot be explained with the KERRBB radiation pattern. The SDSS spectrum (black line) continuum is described with KERRBB (dashed red line, with a shaded blue area representing a confidence interval given by the uncertainty on the spectrum peak; see Sect. \ref{ADemission}) while the torus emission is constrained with the four WISE data points (red dots) and two black bodies (shaded red area with a dashed blue line contour) plotted along with the corresponding temperatures. The thick blue line is the overall model (disk + torus). On each plot, we report the isotropic disk and torus luminosities (in erg/s) and the luminosity ratio $R$ along with its uncertainty (see Sect. \ref{Runcert}). Some archival photometric data (2MASS, NED, GALEX - gray dots) are added to both plots. The yellow shaded area is the luminosity range in which $\nu_{\rm p} L_{\nu_{\rm p}}$ lies, obtained by taking into account different uncertainties (Sect. \ref{Runcert}).} 
\label{plot_example_high}
\end{figure*}

\subsection{Luminosity ratio: total uncertainty} \label{Runcert}

To define an average confidence interval for the main observables used in this work, we chose to neglect possible intrinsic dust absorption (see Sect. \ref{cavo}) and assume no disk variability and an isotropic torus emission. Given that the estimations of all those previous sources of uncertainty (discussed in Sections \ref{cavo} and \ref{cavoT}) are independent and uncorrelated, we defined a confidence interval for $L^{\rm iso}_{\rm d}$, $L^{\rm iso}_{\rm T}$ and $R$ by summing in quadrature the uncertainties coming from the spectrum peak $\nu_{\rm p} L_{\nu_{\rm p}}$, the torus luminosity estimated from WISE data (taking into account the analysis performed with the SPITZER data), the effect of the X-ray corona and the polar and cold star-heated dust on the IR and UV emissions, as discussed in the previous sections. This procedure led to a confidence interval for the disk luminosity of $^{+0.10}_{-0.10}$ dex, and for the torus luminosity of $^{+0.05}_{-0.15}$ dex, resulting in a final average confidence interval for Log $R$ of $^{+0.10}_{-0.20}$ dex (corresponding to a 1$\sigma$ uncertainty).

For illustration, Fig. \ref{plot_example_1} shows the IR--UV SED modeling of one of the SDSS sources (SDSS J103036.93+312028.8, $z=0.8726$): for a fixed viewing angle $\theta_{\rm v} = 30^{\circ}$, the observed luminosity ratio ($R = 1.07^{+0.28}_{-0.39}$) sets a constraint on the BH spin ($a>0.9$) and on the torus aperture angle ($\theta_{\rm T} < 55^{\circ}$). We note that the SS model fails to explain the high observed ratio $R$. For this object, even if all the main sources of uncertainty are taken into account to estimate the interval of $R$, a large BH spin is required to explain the high torus luminosity with respect to that of the disk.

Given the importance of the involved uncertainties in this work, we advise caution when interpreting the confidence interval defined above and the results shown and discussed in the following section: in the worst-case scenario, a direct combination (not in quadrature) of all the previously discussed sources of contamination and their uncertainties could lead to an even larger observed ratio with respect to the intrinsic one; the correction of such a measurement could result in a ratio that is smaller than the one given by the 1$\sigma$ lower limit of its confidence interval by a factor of approximately two, leading to poor or even unavailable estimates of $\theta_{\rm T}$ and $a$.

%%%%%%%%%%%%%%%%%%%%%%%%%%%%%%%%%%%%%%%%%%%%%%%%%%%%%%%%%%%%%%%%%%%%%%%%%%%
%%%%%%%%%%%%%%%%%%%%%%%%%%%%%%%%%%%%%%%%%%%%%%%%%%%%%%%%%%%%%%%%%%%%%%%%%%%
%%%%%%%%%%%%%%%%%%%%%%%%%%%%%%%%%%%%%%%%%%%%%%%%%%%%%%%%%%%%%%%%%%%%%%%%%%%
%%%%%%%%%%%%%%%%%%%%%%%%%%%%%%%%%%%%%%%%%%%%%%%%%%%%%%%%%%%%%%%%%%%%%%%%%%%

\section{Results}\label{results}

In this section, we show the statistical analysis performed on the SDSS sample. As $R$ depends also on the viewing angle of the system (see Sect. \ref{notation}), we fixed it to an average $\theta_{\rm v} = 30^{\circ}$. However, this choice does not influence our results drastically: on average, for a different viewing angle in the range $0^{\circ} < \theta_{\rm v} < 45^{\circ}$, the average BH spin and the torus aperture angle estimates would differ by $< 20 \%$ from our reported results.

%%%%%%%%%%%%%%%%%%%%%%%%%%%%%%%%%%%%%%%%%%%%%%%%%%%%%%%%%%%%%%%%%%%%%%%%%%%

\subsection{Distribution of $R$}

The mean logarithmic value of the luminosity ratio with its standard deviation is $\langle {\rm Log}\ R \rangle = -0.13 \pm 0.20$ (the mean linear value is $\langle R \rangle = 0.83 \pm 0.41$): this value is larger than the SS limit $R = 0.5$ (see Sect. \ref{notation}). The distribution of $R$ is shown in Fig. \ref{ratio_all}: if we take the 1$\sigma$ lower limit of $R$ (given by its average uncertainty), about $60 \%$ of the sources show a ratio that is larger than the SS limit and, for an average torus aperture angle $\theta_{\rm T} = 45^{\circ}$, almost $80 \%$ of the observed ratios cannot be explained with the SS model. Instead, taking into account the relativistic effects related to large BH spin values and the 1$\sigma$ lower limit of $R$ (red histogram), almost $95 \%$ of the luminosity ratios can be explained with the KERRBB model. For the SDSS+HST sample, the mean luminosity ratio does not change significantly with respect to the SDSS sample.

%%%%%%%%%%%%%%%%%%%%%%%%%%%%%%%%%%%%%%%%%%%%%%%%%%%%%%%%%%%%%%%%%%%%%%%%%%%

\subsection{Black hole spin and torus aperture angle}\label{constraints}

The mean values $\langle {\rm Log}\ R \rangle$ and $\langle R \rangle$ (related to the blue histogram in Fig. \ref{ratio_all}) set a constraint on the BH spin: if only the central value is considered, the average BH spin must be $a > 0.95$. If the lowermost limit of $R$ is considered, no constraints on the BH spin can be found (see Fig. \ref{gridpattern2}; see also the previous section). Given that our sample is made of radio-quiet sources, the result of our statistical analysis seems to be in contrast with the idea that all radio-quiet AGNs host slow-rotating BHs (due to the lack of relativistic jets - e.g., \citealt{UrPad}) and in agreement with the assumption that coherent gas accretion (proved by the presence of the BBB in the optical-UV bands; see e.g, \citet{Shang05} and references therein)\footnote{The BBB cannot be modeled properly with other accretion modes, such an advection-dominated accretion flow (ADAF) because for this latter the radiative efficiency is too low to produce a bright emission. Moreover, the ADAF SED is completely different from the AD one (see e.g., \citealt{Nemm}).} through a disk causes the majority of BHs to have large spins (e.g., \citealt{Elvis}; \citealt{Siko} and references therein).

As discussed in \citet{Campiti,Campitib}, an independent BH mass estimate (i.e., virial mass) can be used to constrain $a$ which can be compared with the spin inferred from the luminosity ratios. We performed such an analysis and found no significant correlation between the two spin estimates. There are two main reasons for this lack of correlations: First, different virial equations lead to different BH mass estimates (e.g., the analysis of \citealt{Shenetal11} showed that, with differences up to 1 order of magnitude), and therefore more precise independent mass estimates have to be used to performed such a study. Second, the virial BH mass estimates have a systematic uncertainty of up to $\sim 0.5$ dex (e.g., \citealt{Ves}) which leads to poor spin estimates.

For what concerns the torus aperture angle, neither the SDSS nor the SDSS+HST samples can be used to put significant constraints on $\theta_{\rm T}$, with an average range of $30^{\circ} < \theta_{\rm T} < 70^{\circ}$ (at 2$\sigma$), consistent with the common picture of a torus with an average aperture angle of $45^{\circ}$, and with the results of \citet{Caldero}. For the SS model, the mean luminosity ratios of both samples can be explained only if the average torus aperture angle is smaller than $40^{\circ}$ (at 2$\sigma$), which disagrees with the commonly accepted scenario.

%%%%%%%%%%%%%%%%%%%%%%%%%%%%%%%%%%%%%%%%%%%%%%%%%%%%%%%%%%%%%%%%%%%%%%%%%%%%

\begin{figure*}
\centering
\hskip -0.2 cm
\includegraphics[width=0.51\textwidth]{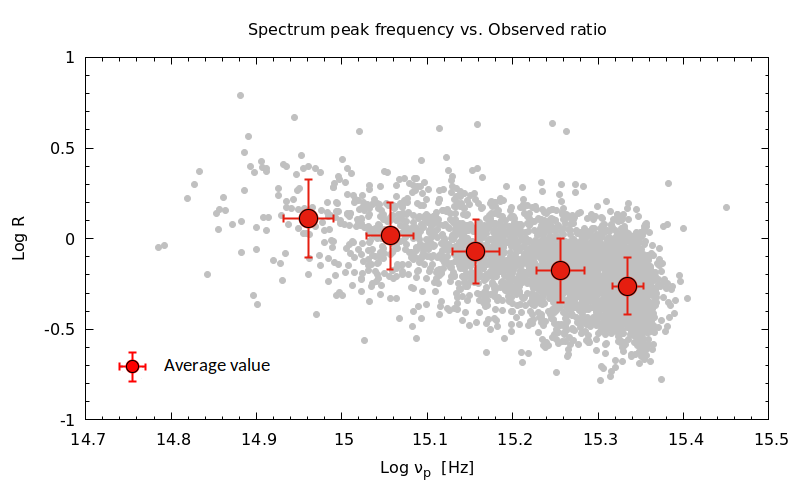}\includegraphics[width=0.43\textwidth]{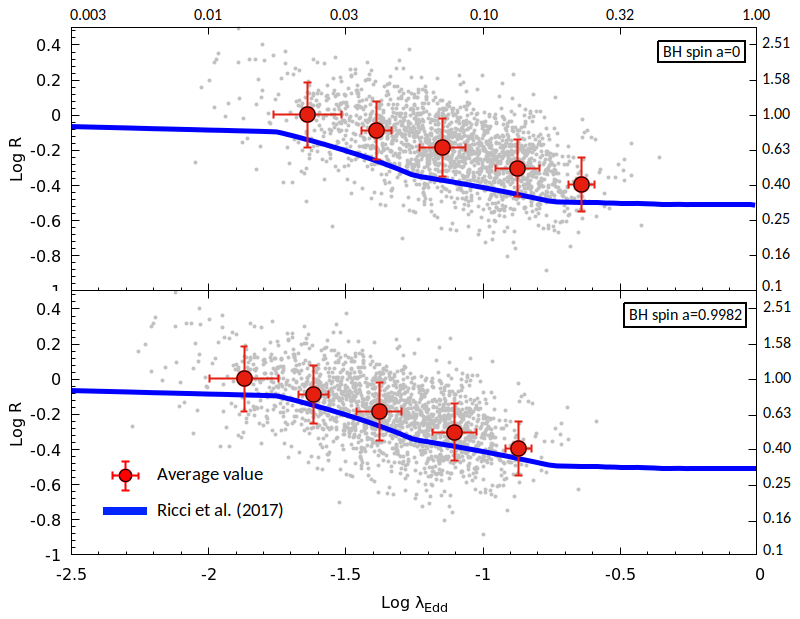}
\caption{Left panel: Luminosity ratio $R$ as a function of the spectrum peak frequency $\nu_{\rm p}$. Right panel: Luminosity ratio $R$ as a function of the Eddington ratio $\lambda_{\rm Edd}$ assuming nonspinning (top panel) and maximally spinning SMBHs (bottom panel) for a fixed $\theta_{\rm v} = 30^{\circ}$ (sharing the same x-axis). The gray dots represent the SDSS sample. The thick blue lines in the right panels represent the results of \citet{Ricci} (their Fig. 4, assuming the covering factor $\approx$ luminosity ratio $R$; see text). On both plots, red dots represent the average values of $R$ computed at fixed frequency and Eddington ratio bins (with 1$\sigma$ error bars). The top x-axis and side y-axis show the linear value of $\lambda_{\rm Edd}$ and $R$, respectively.}
\label{nu_R}
\end{figure*}

\subsection{Sources with large $R$}\label{largeR}

Considering the entire 1$\sigma$ range of $R$ (given by its uncertainty; Sect. \ref{Runcert}), about one-third of the sources of the SDSS sample show a luminosity ratio $R>0.6$ for which the BH spin is constrained to be $a>0.7$ (for a fixed viewing angle $\theta_{\rm v} = 30^{\circ}$; Fig. \ref{ratio_all}). Moreover, for the most extreme values, $\theta_{\rm T}$ must be close to the viewing angle of the system. Almost $5 \%$ of the SDSS sample sources show a ratio at 1$\sigma$ above the KERRBB limit ($\sim 1 \%$ at 2$\sigma$; Fig. \ref{ratio_all}); two such sources are shown as examples in Fig. \ref{plot_example_high}. We checked all those sources to understand the possible causes of those large luminosity ratios: we could not find any characteristic feature to account for these, but data in the mm and FIR bands is lacking (where the peak of the other possible contaminating emissions is located; see Sect. \ref{cavoT}) and our results have to be taken statistically. Dust absorption along the line of sight (e.g. from polar dust) could be one of the possible explanations for those large ratios, along with the contamination of the IR and the absorption of the UV emission, which is grater than average, as discussed and shown in Sect. \ref{cavo} and \ref{cavoT}.

%%%%%%%%%%%%%%%%%%%%%%%%%%%%%%%%%%%%%%%%%%%%%%%%%%%%%%%%%%%%%%%%%%%%%%%%%%%

\subsection{Torus geometry versus Eddington ratio}

Several authors suggested that the key parameter determining the torus covering factor is the Eddington ratio (e.g., \citealt{Law}; \citealt{Ueda}; \citealt{Treister}; \citealt{Burlon}; \citealt{Ezhi}; \citealt{Buch}; \citealt{Ricci}). The $\lambda_{\rm Edd} - R$ anti-correlation (i.e., receding torus) could be due to different factors, as discussed by several authors (e.g., the increase of the inner radius of the obscuring material with incident luminosity, the gravitational potential of the BH, radiative feedback, outflow/inflows, increase of $\dot{M}$: see \citealt{Law}; \citealt{Lama}; \citealt{Menci}; \citealt{Fabianetal09}; \citealt{Wada}; \citealt{Ricci}).

Given the spectral shape similarity between KERRBB models and the degeneracy between the parameters ($a$, $M$, $\dot{M}$), once the BH spin is fixed, it is possible to constrain the BH mass and the accretion rate and hence $\lambda_{\rm Edd}$. However, in this study, both $\lambda_{\rm Edd}$ and $R$ depend on the same variable (i.e. the peak luminosity $\nu_{\rm p} L_{\nu_{\rm p}}$) even though in a different manner ($\sqrt{\nu_{\rm p} L_{\nu_{\rm p}}}$ vs. $1/\nu_{\rm p} L_{\nu_{\rm p}}$, respectively; see \citealt{Campitib} and Sect. \ref{notation}). We chose instead to study the relation between the peak frequency $\nu_{\rm p}$ and $R$, because these are independent measurements: we find an anti-correlation between the two quantities as shown in the left panel of Fig. \ref{nu_R} (the best fit is Log $R \propto -1.01\ {\rm Log}\ \nu_{\rm p} + 15.28$, with a 1$\sigma$ data dispersion of $\sim 0.15$ dex). Assuming that no strong dust absorption is present in those sources (causing an apparent shift of the peak frequency to lower values; see the discussion in Sect. \ref{cavo}), from this plot, two simple conclusions can be drawn:
\begin{itemize}
	\item Large values of $\nu_{\rm p}$ are related to small BH masses (given the dependence $M \propto 1/\nu^{2}$; see \citealt{Campiti,Campitib}), and therefore large Eddington ratios are expected for those sources,\footnote{For our sample, we find that the proportionality $\lambda_{\rm Edd} \propto M^{-1}$ is valid with a dispersion of $\sim 0.2$ dex at 1$\sigma$.} linked to small observed ratios.
	\item On the contrary, small values of $\nu_{\rm p}$ are related to large BH masses, and therefore large values of $R$ correspond to small Eddington ratios.
\end{itemize}

These conclusions are in agreement with the receding torus scenario and with the work of \citet{Ricci} who studied a local sample of AGNs (median redshift $z \approx 0.037$). In order to visualize these results, we plotted the observed ratios as a function of $\lambda_{\rm Edd}$ (right panel of Fig. \ref{nu_R}) for two fixed BH spins ($a=0 - 0.9982$) and $\theta_{\rm v} = 30^{\circ}$, although, as discussed before, the two quantities are biased by the spectrum peak luminosity.\footnote{For different values of $\theta_{\rm v}$ in the range $0^{\circ} < \theta_{\rm v} < 45^{\circ}$, the estimated $\lambda_{\rm Edd}$ would differ by a factor of $<20 \%$ from the reported ones.} We noticed that the anti-correlation discussed above is visible and comparable with the results of \citet{Ricci} (blue lines). For such a comparison, we removed the effect of a possible X-ray corona above the disk which can modify $L^{\rm iso}_{\rm d}$ and therefore $\lambda_{\rm Edd}$ and $R$ (following the work of \citet{Duras} and the discussion in Sect. \ref{cavo}) and assuming that the covering factor shown by Ricci et al. in their Figure 4 corresponds to the luminosity ratio $R$.\footnote{Even though the two quantities do not necessarily coincide, the thick blue lines in Fig. \ref{nu_R} (right panel) must be considered only as an indicative representation of the relationship between them and not a general function.}

\begin{figure*}
\centering
\hskip -0.2 cm
\includegraphics[width=0.95\textwidth]{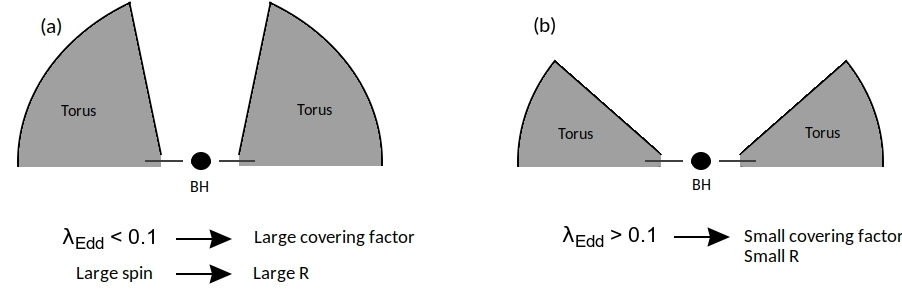}
\caption{Schematic cartoons representing the different results inferred in our work. Despite the value of the BH spin, if the Eddington ratio is low ($\lambda_{\rm Edd} < 0.1$), the torus can survive in almost all directions resulting in a large covering factor (\textit{a}); on the contrary, for larger $\lambda_{\rm Edd}$, the torus only survives on the equatorial plane with a small covering factor and luminosity ratio $R$ (\textit{b}). For large ratios ($R \sim 1$), the covering factor must be large and intercepts almost all the radiation coming from the AD. Moreover, large $R$ values can be explained only if the BH is rapidly spinning (Sect. \ref{notation}).} 
\label{cartoon_finale}
\end{figure*}

Although our estimates lie above the results of Ricci et al., there is a clear and remarkable resemblance between their slope and ours in the same Eddington ratio bin, with an even more evident compatibility between the two works when focusing on maximally spinning SMBHs. Moreover, the recent results of \citet{Toba} are also partially compatible with ours, although the anti-correlation found by these latter authors (shown in their Figure 7, not shown in our Fig. \ref{nu_R} for clarity) is "flatter" than ours and the one from Ricci et al. The discrepancy between our results and those of \citet{Ricci} could be caused by (1) the different luminosity ranges of the two samples, (2) the different redshift ranges related to the sources in the two works, and (3) the fact that the high-$z$ sources show a larger $\lambda_{\rm Edd}$ (e.g., \citealt{Lusso}).

As discussed above, our results suggest that large Eddington ratios result in small luminosity ratios and, for a fixed spin value, in small covering factors (i.e., large $\theta_{\rm T}$): the torus only survives on the equatorial plane. On the contrary, smaller Eddington ratios could result in a large $R$ and large covering factors: the AGN is not able to remove dust in all directions efficiently which forms a more extended structure with a small $\theta_{\rm T}$ (Fig. \ref{cartoon_finale} with an indicative threshold between the two scenarios at $\lambda_{\rm Edd} \sim 0.1$), in agreement with the scenario proposed by \citet{Ricci} (see also \citealt{Kawaka}; \citealt{Lyu2017}). These results are in overall agreement with the suggestions by \citet{Ishi} (see also \citealt{Ishi20}): these latter authors discuss the possibility that the radiation pattern of the AD shapes the surrounding structures, depending on the BH spin. For low spin values, dust can be cleared out in the face-on direction while it may survive at higher inclination angles; for high spin values, dust can be removed from most directions, except in the equatorial plane (resulting in a smaller covering factor with respect the low-spin case). However, for approximately one-third of our sample, the large luminosity ratios can be explained only if both the torus covering factor and the BH spin are large (see Sect. \ref{largeR}).

%%%%%%%%%%%%%%%%%%%%%%%%%%%%%%%%%%%%%%%%%%%%%%%%%%%%%%%%%%%%%%%%%%%%%%%%%%%

\subsection{Hot dust covering factor}

Dust distribution can play an important role in describing the IR emission properly. \citet{Honig} reconsiders the torus structure and emission based on observational constraints: the author states that the IR emission in different bands is due to different regions of the dust distribution, in particular, the NIR emission is due to a disk-like structure with a covering factor of $\sim 0.2 - 0.3$ (see also \citealt{MorNet} and \citealt{Landt} who found an average covering factor of $\sim 0.1$), corresponding to $\theta_{\rm T} \sim 70^{\circ} - 80^{\circ}$. Given that the NIR torus luminosity estimate is free from the possible contamination related to the cold and polar dust (see Sect. \ref{cavoT} and Fig. \ref{colddust}), we used the hot black-body emission ($L^{\rm iso}_{\rm T, hot}$) as a proxy for the NIR luminosity to compute the NIR luminosity ratio, defined as $R_{\rm NIR} = L^{\rm iso}_{\rm T, hot} / L^{\rm iso}_{\rm d}$: for the SDSS sample, we found $\langle {\rm Log}\ R_{\rm NIR} \rangle = -0.48 \pm 0.20$ (the mean linear value is $\langle R_{\rm NIR} \rangle = 0.37 \pm 0.18$)\footnote{For the SDSS+HST sample we obtained $\langle {\rm Log}\ R_{\rm NIR} \rangle = -0.58 \pm 0.22$ (the linear mean is $\langle R_{\rm NIR} \rangle = 0.29 \pm 0.15$).} This result is only consistent with the one described by H\"{o}nig if the BH spin is $a>0.95$ at 1$\sigma$ ($a>0.5$ at 2$\sigma$). Using the SS model, the NIR luminosity ratio constrains the torus aperture angle to within a range of $\theta_{\rm T} < 60^{\circ}$ at 2$\sigma$, which is inconsistent with the average NIR covering factor.

%%%%%%%%%%%%%%%%%%%%%%%%%%%%%%%%%%%%%%%%%%%%%%%%%%%%%%%%%%%%%%%%%%%%%%%%%%%
%%%%%%%%%%%%%%%%%%%%%%%%%%%%%%%%%%%%%%%%%%%%%%%%%%%%%%%%%%%%%%%%%%%%%%%%%%%
%%%%%%%%%%%%%%%%%%%%%%%%%%%%%%%%%%%%%%%%%%%%%%%%%%%%%%%%%%%%%%%%%%%%%%%%%%%
%%%%%%%%%%%%%%%%%%%%%%%%%%%%%%%%%%%%%%%%%%%%%%%%%%%%%%%%%%%%%%%%%%%%%%%%%%%

\section{Summary, Discussion, and Conclusions} \label{sec-concl} 

We selected a sample of approximately $2000$ Type 1 AGNs from the SDSS DR7Q catalog (\citealt{Shenetal11}) with a redshift of $0.35<z<0.89$ and a bolometric luminosity (computed with the corrections from \citealt{Rich06}) of grater than $10^{46}$ erg/s. We studied the distribution of the luminosity ratio $R$ between the dusty torus luminosity $L^{\rm iso}_{\rm T}$ and the AD one $L^{\rm iso}_{\rm d}$ inferred from the observed SEDs in order to constrain the torus aperture angle $\theta_{\rm T}$ and possibly the adimensional BH spin $a$. Our statistical analysis and results are summarized below:
\begin{itemize}
    \item We used two black bodies to infer $L^{\rm iso}_{\rm T}$ from the WISE data with a temperature of $T < 2000$ K and assumed that the torus spatial distribution is described by its aperture angle $\theta_{\rm T}$. The possible clumpiness and anisotropy of the torus dust distribution might have an important effect on $L^{\rm iso}_{\rm T}$ (discussed in Sect. \ref{cavoT}). Given the limited spectral coverage, numerical modelings with a clumpy dust distribution are hard to apply, we therefore assume the dust distribution to be continuous and the torus emission to be isotropic. The BBB in the optical-UV range is described by the relativistic AD model KERRBB.

    \item To evaluate the uncertainties related to our IR and UV fitting procedure, we built two subsamples by cross-matching our sample with the HST and SPITZER catalogs: for a few sources, the collected spectroscopic data in the FUV and IR bands allowed us to obtain a better constraint on both the disk and torus emissions. On average, $L^{\rm iso}_{\rm d}$ inferred with the SDSS spectrum alone is underestimated by a factor of $\sim 0.05$ dex, while the two black bodies approximation for the IR emission proved to be reliable.

    \item We analyzed and quantified the uncertainties related to the main possible sources of contamination and absorption of the torus and disk luminosities, along with the effect of the presence of polar dust, an X-ray corona above the disk and the possible anisotropy of the torus emission. In the case of an isotropic torus emission, we estimated an average confidence interval for the luminosity ratio $R$ of $^{+0.10}_{-0.20}$ dex by summing the involved uncertainties in quadrature. However, if these latter are directly combined (not in quadrature), the lower limit on the ratio confidence interval could be smaller by a factor of about two, leading to poor estimates of $\theta_{\rm T}$ and $a$.
    
    \item The mean logarithmic value of the luminosity ratio is $\langle {\rm Log}\ R \rangle = -0.13 \pm 0.20$ (the linear mean is $\langle R \rangle = 0.83 \pm 0.41$). Using the KERRBB relativistic AD angular pattern, we set constraints on the average torus aperture angle and the BH spin: assuming a viewing angle $\theta_{\rm v} = 30^{\circ}$, the average ratio corresponds to a torus aperture angle in the range $30^{\circ} < \theta_{\rm T} < 70^{\circ}$, and an average BH spin of $a \sim 0.95$ (if only the central value of $R$ is considered). If the lowermost limit of $R$ is considered, no constraints on the BH spin can be found.
    
    \item Even though all the main sources of contamination of the IR and UV luminosities are taken into account, about one-third of the sources show a large value of $R$ that can be explained by very strong relativistic effects, that is, the SMBH is rapidly spinning with $a>0.7$. The same conclusion can be drawn when using the hot black-body emission as a proxy for the NIR luminosity, which is thought to be due to a disk-like structure with a covering factor of $\sim 0.2 - 0.3$ (\citealt{Honig}).
    
	\item Although our sample has been built choosing only radio-quiet sources, our statistical results suggest that a fraction of the sources might host a rapidly spinning SMBH, in contrast with the view of slow-rotating BHs at the center of those AGNs. Although our findings clearly require further investigation in order to assess their robustness, their implications are as follows:
	\begin{itemize}
		\item Despite their radio nature, the accretion mode is the same for the majority of AGNs, i.e., a coherent gas accretion that spins the BHs up to the maximum value (following the accretion disk theory of e.g., \citealt{Thorn74}).
		\item If relativistic jets are indeed linked to rapidly spinning SMBHs, the radio-quiet nature of some sources could be due to the different inclination angle of the system or to the jet dissipation region (see e.g., \citealt{GhiTav15}).
		\item If relativistic jets are present only in radio-loud sources, their production could be linked to some other features of the system and not only to the BH spin.\\
	\end{itemize} 	
	
	\item The evolution of the hole could play an important role for both the spin and the shape of the surrounding environment. Several studies suggest that the slower spinning SMBHs should be the most massive ones because a lower radiative efficiency (linked to a small value of $a$) favors a faster growth of the BH, regardless of the nature of the accretion (chaotic vs. coherent; see e.g. \citealt{Campitib}; \citealt{Zubo} and references therein). Although it is extrapolated from a very small sample, such a conclusion is supported by the available spin measurements (see e.g. \citealt{Rey} and references therein). In this work, we find no significant correlation between the BH mass (virial and from the fitting procedure) and the observed ratio (or high BH spin values constrained from $R$). Given the involved uncertainties for both the mass and the spin estimates, we were not able to carry out a proper comparison with the referred studies and therefore cannot draw any robust conclusions. However given the average virial BH mass of the sample (Log $M/M_{\odot} \sim 9$), our results, which indicate the presence of highly spinning SMBHs, agree with the work of \citet{Trak}, who, in contrast to some of studies mentioned above, found that very massive BHs have large radiative efficiencies (i.e., large spins). Given that these arguments are still a matter of debate, we suggest that caution be taken when interpreting the conclusions presented here because different works show different results regarding the possible relation between $M$ and $a$.
	
	\item Following this last point, it is well known that the accretion history of the BH and its accretion rate are crucial to understanding its evolution and the link between the different features: in this context, other variables must be taken into account that are not completely understood and/or are very hard to constrain (e.g., the exact BH accretion mode evolution --coherent or chaotic--, the BH system --isolated or binary--, and so on). The presence of the BBB in the optical-UV bands for the sources analyzed in this work suggests that the accretion mode is coherent and can be described by an AD around the SMBH. Despite this, and the evolution of the spin in such a scenario (which lead it to the maximum value $a \sim 1$; \citealt{Thorn74}), the growth of the BH is linked to $\dot{M}$ and its possible changes during its accretion history (e.g., via gas or BH--BH mergers); this latter can be studied only for only very few sources at high redshift, which cannot be used to understand the overall evolution of AGNs through time. Given all these considerations, some general conclusions can be drawn:
	\begin{itemize}
		\item If accretion occurs in a chaotic way, the BH spin is expected to be small or even approximately zero, and in this scenario the BH can gather mass very efficiently. If a coherent accretion phase follows the fast BH growth, the BH can spin up to the maximum value when the hole doubles its mass (\citealt{Thorn74}) roughly after $\sim 0.1$ Gyr (using the definition of Salpeter time, with an average radiative efficiency of $\eta \sim 0.1$ and an Eddington ratio of $\lsim 1$; \citealt{Salp}; see also \citealt{Campitib}).
		\item If a BH grows its mass via mergers, then its spin depends on the initial spins of the two colliding holes. As in the previous point, if gas is present around the system, regardless of the value of the initial BH spin, this latter can increase up to the maximum value through a possible coherent accretion mode in a short amount of time.
		\item If the BH growth occurs via an AD, the hole will increase its mass less efficiently than the previous cases but its spin will very quickly become large.\\
	\end{itemize} 
	
	Despite these general conclusions and given the uncertainties involved in such studies, these latter arguments are still too weak to draw strong conclusions. Therefore, we advise caution when interpreting the relation discussed in the previous point: given that both the BH mass and its spin depend on the evolution of the hole through time, their possible (and possibly general) relation can only be studied with a very large sample of data.

	\item Our results suggest an anti-correlation between the luminosity ratio $R$ and the Eddington ratio $\lambda_{\rm Edd}$, as also suggested by several authors (e.g., \citealt{Ezhi}; \citealt{Buch}) and comparable with the results of \citet{Ricci} (Fig. \ref{nu_R}, right panel). A larger $\lambda_{\rm Edd}$ leads to smaller ratios $R$ and, for a fixed spin value, to smaller covering factors (i.e., the torus survives only on the equatorial plane). On the contrary, a smaller Eddington ratio results in a larger $R$ and a larger covering factor (partially in agreement with the works by \citealt{Ishi} and \citealt{Ishi20}).
\end{itemize}

The statistical analysis and the results presented in this paper suggest the presence of spinning SMBHs surrounded by a dusty torus whose structure depends on the AD luminosity. The ongoing improvement of numerical models for fitting multi-frequency SEDs and comparisons with different approaches to constrain the AGN parameters are both needed to verify our findings and to strengthen the use of the relativistic AD pattern to study the physics of BHs.\\

%%%%%%%%%%%%%%%%%%%%%%%%%%%%%%%%%%%%%%%%%%%%%%%%%%%%%%%%%%%%%%%%%%%%%%%%%%%
%%%%%%%%%%%%%%%%%%%%%%%%%%%%%%%%%%%%%%%%%%%%%%%%%%%%%%%%%%%%%%%%%%%%%%%%%%%
%%%%%%%%%%%%%%%%%%%%%%%%%%%%%%%%%%%%%%%%%%%%%%%%%%%%%%%%%%%%%%%%%%%%%%%%%%%
%%%%%%%%%%%%%%%%%%%%%%%%%%%%%%%%%%%%%%%%%%%%%%%%%%%%%%%%%%%%%%%%%%%%%%%%%%%

\begin{acknowledgements}
We are grateful to the anonymous referee for her/his constructive critical comments and suggestions, which helped improving the paper.
\end{acknowledgements}

%%%%%%%%%%%%%%%%%%%%%%%%%%%%%%%%%%%%%%%%%%%%%%%%%%%%%%%%%%%%%%%%%%%%%%%%%%%
%%%%%%%%%%%%%%%%%%%%%%%%%%%%%%%%%%%%%%%%%%%%%%%%%%%%%%%%%%%%%%%%%%%%%%%%%%%
%%%%%%%%%%%%%%%%%%%%%%%%%%%%%%%%%%%%%%%%%%%%%%%%%%%%%%%%%%%%%%%%%%%%%%%%%%%
%%%%%%%%%%%%%%%%%%%%%%%%%%%%%%%%%%%%%%%%%%%%%%%%%%%%%%%%%%%%%%%%%%%%%%%%%%%

\medskip
\let\itshape\upshape

\begingroup
\let\clearpage\relax

\label{lastpage}
\endgroup


\begin{thebibliography}{}

\bibitem[Alonso-Herrero et al.(2011)]{Alonso}
Alonso-Herrero, A., Ramos Almeida, C., Mason, R. et al., 2011,
\newblock \emph{ApJ, 736, 82}

\bibitem[Antonucci(1993)]{Anto}
Antonucci, R., 1993,
\newblock \emph{ARA\&A, 31, 473}

\bibitem[Arnaud(1996)]{ArnaudXPS}
Arnaud, K. A., 1996,
\newblock \emph{Astronomical Data Analysis Software and Systems V. ASP Conference Series, Vol. 101, eds. G. H. Jacoby and J. Barnes, p. 17.}

\bibitem[Asmus et al.(2016)]{Asmus16}
Asmus D., H\"{o}nig S. F. \& Gandhi P., 2016,
\newblock \emph{ApJ, 822, 109}

\bibitem[Asmus (2019)]{Asmus19}
Asmus D., 2019,
\newblock \emph{MNRAS, 489(2), 2177-2188}

\bibitem[Baron et al.(2016)]{Baron}
Baron, D., Stern, J., Poznanski, D. \& Netzer, H., 2016,
\newblock \emph{APJ, 832.1, 8}

\bibitem[Barvainis(1987)]{Barva}
Barvainis, R., 1987,
\newblock \emph{ApJ, 320, 537}

\bibitem[Becker et al.(1995)]{Becker}
Becker, R. H., White, R. L. \& Helfand, D. J., 1995, 
\newblock \emph{ApJ, 450, 559}.

\bibitem[Bendo et al.(2003)]{Bendo}
Bendo, G. J., Joseph, R. D., Wells, M. et al., 2003,
\newblock \emph{ApJ, 125, 2361}

\bibitem[Boselli et al.(2010)]{Bose}
Boselli, A., Ciesla, L., Buat, V. et al., 2010,
\newblock \emph{A\&A, 518, L61}

\bibitem[Buchner \& Bauer(2017)]{Buch}
Buchner, J. \& Bauer, F. E., 2017,
\newblock \emph{MNRAS, 465:434-4362}

\bibitem[Burlon et al.(2011)]{Burlon}
Burlon, D., Ajello, M., Greiner, J. et al., 2011,
\newblock \emph{ApJ, 728, 58}

\bibitem[Calderone et al.(2012)]{Caldero12}
Calderone, G., Sbarrato, T., \& Ghisellini, G., 2012,
\newblock \emph{MNRAS, 425, L41}

\bibitem[Calderone et al.(2013)]{Caldero}
Calderone, G., Ghisellini, G., Colpi, M., \& Dotti, M., 2013,
\newblock \emph{MNRAS, 431, 210}

\bibitem[Calzetti et al.(2000)]{Calze}
Calzetti, D., Armus, L., Bohlin, R. C., Kinney, A. L., Koornneef, J. \& Storchi-Bergmann, T., 2000,
\newblock \emph{ApJ, 533(2), 682}

\bibitem[Campitiello et al.(2018)]{Campiti}
Campitiello, S., Ghisellini, G., Sbarrato, T., \& Calderone, G., 2018,
\newblock \emph{A\&A, 612, A59}

\bibitem[Campitiello et al.(2019)]{Campitib}
Campitiello, S., Celotti, A., Ghisellini, G. \& Sbarrato, T., 2019,
\newblock \emph{A\&A, 625, A23}

\bibitem[Campitiello et al.(2020)]{Campitic}
Campitiello, S., Celotti, A., Ghisellini, G. \& Sbarrato, T., 2020,
\newblock \emph{A\&A, 640, A39}

\bibitem[Cardelli et al.(1989)]{Cardelli}
Cardelli, J., A., Clayton, G., C. \& Mathis, J., S., 1989,
\newblock \emph{ApJ, 345, 245}

\bibitem[Castignani et al.(2013)]{Castietal}
Castignani, G., Haardt, F., Lapi, A., et al., 2013,
\newblock \emph{A\&A, 560, 28}

\bibitem[Castignani \& De Zotti(2015)]{CastiDe}
Castignani, G., \& De Zotti, G., 2015,
\newblock \emph{A\&A, 573, A125}

\bibitem[Chainakun et al.(2019)]{Chaina}
Chainakun P., Watcharangkool A., Young A. J. \& Hancock S., 2019,
\newblock \emph{MNRAS, 487.1, 667-680}

\bibitem[Collinson et al.(2016)]{Collinson}
Collinson, J. S., Ward, M. J., Landt, H. et al., 2016,
\newblock \emph{MNRAS, stw2666}

\bibitem[Cunningham(1975)]{Cunnin}
Cunningham, C.~T., 1975,
\newblock \emph{ApJ, 202, 788}

\bibitem[Dale et al.(2012)]{Dale}
Dale, D. A., Aniano, G., Engelbracht, C. W. et al., 2012,
\newblock \emph{ApJ, 745, 95} 

\bibitem[Davis \& Laor(2011)]{DavLao}
Davis, S., W. \& Laor, A., 2011,
\newblock \emph{ApJ, 728, 98}

\bibitem[Deo et al.(2011)]{Deoetal}
Deo R. P., Richards G. T., Nikutta R. et al., 2011,
\newblock \emph{ApJ, 729, 108}

\bibitem[Done et al.(2012)]{Done}
Done, C., Davis, S. W., Jin, C., Blaes, O. \& Ward, M., 2012,
\newblock \emph{MNRAS, 420, 1848}

\bibitem[Duras et al.(2020)]{Duras}
Duras, F., Bongiorno, A., Ricci, F. et al., 2020,
\newblock \emph{A\&A, 636, A73}

\bibitem[Elvis et al.(2002)]{Elvis}
Elvis M., Risaliti G. \& Zamorani G., 2002,
\newblock \emph{ApJL, 565.2, L75}

\bibitem[Ezhikode et al.(2017)]{Ezhi}
Ezhikode S. H., Gandhi P., Done C. et al., 2017,
\newblock \emph{MNRAS, 472, 3492}

\bibitem[Fabian et al.(2009)]{Fabianetal09}
Fabian A., Vasudevan R., Mushotzky R. et al., 2009,
\newblock \emph{MNRAS, 394, L89-L92}

\bibitem[Fritz et al.(2006)]{Fritz}
Fritz, J., Franceschini, A., \& Hatziminaoglou, E., 2006,
\newblock \emph{MNRAS, 366, 767}

\bibitem[Gallagher et al.(2006)]{Galla}
Gallagher, S. C., Brandt, W. N., Chartas, G. et al., 2006, 
\newblock \emph{ApJ, 644, 709}

\bibitem[Ghisellini, Haardt \& Matt(1994)]{GhiHa}
Ghisellini G., Haardt F. \& Matt G., 1994,
\newblock \emph{MNRAS, 267, 743}

\bibitem[Ghisellini \& Tavecchio (2015)]{GhiTav15}
Ghisellini, G. \& Tavecchio, F., 2015,
\newblock \emph{MNRAS, 448.2: 1060-1077}

\bibitem[Granato \& Danese(1994)]{GraDan}
Granato, G. L., \& Danese, L., 1994, 
\newblock \emph{MNRAS, 268, 235}

\bibitem[Gu(2013)]{Gu}
Gu, M., 2013,
\newblock \emph{ApJ, 773, 176}

\bibitem[Hao et al.(2013)]{Hao}
Hao, H., Elvis, M., Bongiorno, A. et al., 2013,
\newblock \emph{MNRAS, 434, 3104}

\bibitem[Haardt \& Madau(2012)]{Haamad}
Haardt, F., \& Madau, P., 2012,
\newblock \emph{ApJ, 746, 125}

\bibitem[Hernán-Caballero et al.(2004)]{Hernan}
Hernán-Caballero, A., Hatziminaoglou, E., Alonso-Herrero,  A. \& Mateos, S., 2016,
\newblock \emph{MNRAS, 463.2, 2064-2078}

\bibitem[H\"{o}nig(2019)]{Honig}
H\"{o}nig, S. F., 2019,
\newblock \emph{ApJ, 884.2, 171}

\bibitem[H\"{o}nig \& Kishimoto(2010)]{HonKish}
H\"{o}nig, S. F., \& Kishimoto, M., 2010,
\newblock \emph{A\&A, 523, 27}

\bibitem[H\"{o}nig \& Kishimoto(2017)]{HonKish17}
H\"{o}nig, S. F., \& Kishimoto, M., 2017,
\newblock \emph{ApJL, 838(2), L20}

\bibitem[Hubeny et al.(2000)]{Hub2000}
Hubeny, I., Agol, E., Blaes, O., \& Krolik, J. H., 2000,
\newblock \emph{ApJ, 533, 710}

\bibitem[Ishibashi(2020)]{Ishi20}
Ishibashi, W., 2020,
\newblock \emph{MNRAS, 495(2), 2515-2523}

\bibitem[Ishibashi et al.(2019)]{Ishi}
Ishibashi, W., Fabian, A. \& Reynolds, C., 2019,
\newblock \emph{MNRAS, 486.2, 2210-2214}

\bibitem[Jaffe et al.(2004)]{Jaffe}
Jaffe, W., Meisenheimer, K., Röttgering, H. J. A., et al., 2004,
\newblock \emph{Nature, 429, 47}

\bibitem[Kawakatu \& Ohsuga(2011)]{Kawaka}
Kawakatu, N. \& Ohsuga, K., 2011,
\newblock \emph{MNRAS, 417.4, 2562-2570}

\bibitem[Kellermann et al.(1989)]{Keller}
Kellermann, K. I., Sramek, R., Schmidt, M. et al., 1989,
\newblock \emph{ApJ, 98, 1195-1207}

\bibitem[Kennicutt et al.(2003)]{Kennetal}
Kennicutt Jr, R. C., Armus, L., Bendo, G. et al.  2003,
\newblock \emph{PASP, 115(810), 928}

\bibitem[Koratkar \& Blaes(1999)]{Kora}
Koratkar, A. \& Blaes, O., 1999,
\newblock \emph{PASP, 111, 1}

\bibitem[Krogager et al.(2015)]{Krog}
Krogager, J., Geier, S., Fynbo, J. et al., 2015,
\newblock \emph{ApJ Supplement Series, 217.1, 5}

\bibitem[Krolik \& Begelman(1988)]{KroBeg}
Krolik, J. H., \& Begelman, M. C., 1988,
\newblock \emph{ApJ, 329, 702}

\bibitem[Kubota \& Done(2018)]{KuDo}
Kubota, A., \& Done, C., 2018, 
\newblock \emph{MNRAS, 480.1, 1247-1262}

\bibitem[Lamastra et al.(2006)]{Lama}
Lamastra A., Perola G. C. \& Matt G., 2006,
\newblock \emph{A\&A, 449, 551-558}

\bibitem[Landt et al.(2011)]{Landt}
Landt, H. et al., 2011,
\newblock \emph{MNRAS, 414, 218-240}

\bibitem[Laor \& Netzer(1989)]{LaoNet}
Laor, A. \& Netzer, H., 1989,
\newblock \emph{MNRAS 238, 897}

\bibitem[Lawrence(1991)]{Law}
Lawrence, A., 1991, 
\newblock \emph{MNRAS, 252, 586-592}

\bibitem[Lawrence \& Elvis(1982)]{LawElv}
Lawrence, A. \& Elvis, M., 1982, 
\newblock \emph{ApJ, 256, 410-426}

\bibitem[Leftley et al.(2018)]{Left}
Leftley J. H., Tristram K. R. W., H\"{o}nig S. F. et al., 2018,
\newblock \emph{ApJ, 862, 17}

\bibitem[Leipski et al.(2014)]{Leip}
Leipski, C., Meisenheimer, K., Walter, F. et al., 2014,
\newblock \emph{ApJ, 785.2, 154}

\bibitem[Li et al.(2005)]{Lietal}
Li, L.-X., Zimmerman, E.~R., Narayan, R., \& McClintock, J.~E., 2005,
\newblock \emph{ApJ, 157, 335-370}

\bibitem[López-Gonzaga et al.(2016)]{Lopez}
López-Gonzaga N., Burtscher L., Tristram K. R. W., Meisenheimer K. \& Schartmann M., 2016,
\newblock \emph{A\&A, 591, A47}

\bibitem[Lusso \& Risaliti(2017)]{LuRi}
Lusso E. \& Risaliti G., 2017, 
\newblock \emph{A\& A, 602, A79}

\bibitem[Lusso et al.(2012)]{Lusso}
Lusso, E., Comastri, A., Simmons, B. D. et al., 2012,
\newblock \emph{MNRAS, 425, 623}

\bibitem[Lyu, Rieke \& Shi(2017)]{Lyu2017}
Lyu, J., Rieke, G. H. \& Shi, Y., 2017,
\newblock \emph{ApJ, 835, 257}

\bibitem[Lyu \& Rieke(2018)]{Lyu}
Lyu, J. \& Rieke, G. H., 2018,
\newblock \emph{ApJ, 866.2, 92}

\bibitem[Lyu, Rieke \& Smith (2019)]{Lyu19}
Lyu, J., Rieke, G. H. \& Smith, P. S., 2019,
\newblock \emph{ApJ, 866.1, 33}

\bibitem[Madau(1995)]{Madau}
Madau, P, 1995,
\newblock \emph{ApJ, 441, 18-27}

\bibitem[Manucci et al.(2001)]{Manuc}
Mannucci, F., Basile, F., Poggianti, B. M. et al., 2001,
\newblock \emph{MNRAS, 326, 745}

\bibitem[Ma \& Wang(2013)]{MaWa}
Ma, X.-C., \& Wang, T.-G., 2013, 
\newblock \emph{MNRAS, 430, 3445}

\bibitem[McClintock et al.(2006)]{McClint}
McClintock, J. E., Shafee, R., Narayan, R., et al., 2006,
\newblock \emph{ApJ, 652, 518}

\bibitem[Menci et al.(2008)]{Menci}
Menci N., Fiore F., Puccetti S. \& Cavaliere A., 2008,
\newblock \emph{ApJ, 686, 219-229}

\bibitem[Merloni et al.(2014)]{Merloni}
Merloni, A., Bongiorno, A., Brusa, M. et al., (2014),
\newblock \emph{MNRAS, 437, 3550}

\bibitem[Miniutti \& Fabian(2004)]{Miniu}
Miniutti, G. \& Fabian, A. C., 2004,
\newblock \emph{MNRAS, 349, 1435}

\bibitem[Mor et al.(2009)]{Mor}
Mor, R., Netzer, H., \& Elitzur, M. 2009,
\newblock \emph{ApJ, 705, 298}

\bibitem[Mor \& Netzer(2012)]{MorNet}
Mor R. \& Netzer H., 2012,
\newblock \emph{MNRAS, 420, 526}

\bibitem[Nemmen et al. (2008)]{Nemm}
Nemmen, R. S., et al., 2014,
\newblock \emph{MNRAS, 438(4), 2804-2827}

\bibitem[Nenkova et al.(2008a)]{Nenka}
Nenkova, M., Sirocky, M. M., Ivezić, Z., \& Elitzur, M. 2008a,
\newblock \emph{ApJ, 685, 147}

\bibitem[Nenkova et al.(2008b)]{Nenkb}
Nenkova, M., Sirocky, M. M., Nikutta, R. et al., 2008b,
\newblock \emph{ApJ, 685, 160}

\bibitem[Neugebauer et al.(1979)]{Neugetal}
Neugebauer, G., Oke, J.~B., Becklin, E.~E., \& Matthews, K., 1979,
\newblock \emph{ApJ, 230, 79}

\bibitem[Pearson et al.(2013)]{Pear}
Pearson, E. A., Eales, S., Dunne, L., et al., 2013,
\newblock \emph{MNRAS, 435.4, 2753-2763}

\bibitem[Petrucci et al.(2017)]{Petru}
Petrucci P. O., Ursini F., De Rosa A. et al., 2018,
\newblock \emph{A\&A, 611, A59}

\bibitem[Pier \& Krolik(1993)]{PierKro}
Pier, E. A., \& Krolik, J. H., 1993,
\newblock \emph{ApJ, 418, 673}

\bibitem[Rees et al.(1969)]{Reesetal}
Rees, M. J., Silk, J., Werner, M. et al., 1969,
\newblock \emph{Nature, 223, 788}

\bibitem[Reynolds (2020)]{Rey}
Reynolds, C. S., 2020,
\newblock \emph{Annual Review of A\&A, 59}

\bibitem[Ricci et al.(2017)]{Ricci}
Ricci C., Trakhtenbrot B., Koss M. J. et al., 2017,
\newblock \emph{Nature, 549(7673), 488-491}

\bibitem[Richards et al.(2006)]{Rich06}
Richards, G., T., Lacy, M., Storrie-Lombardi, L. J. et al., 2006,
\newblock \emph{ApJS, 166, 470}

\bibitem[Risaliti, Elvis \& Nicastro(2002)]{Risa}
Risaliti, G., Elvis, M. \& Nicastro, F., 2002,
\newblock \emph{ApJ, 571.1, 234}

\bibitem[Sadowski(2009)]{Sad09} 
Sadowski A., 2009,
\newblock \emph{ApJS, 183, 171}

\bibitem[Sadowski et al.(2009)]{SadwAbra09} 
Sadowski A., Abramowicz M., Bursa M. et al., 2009,
\newblock \emph{A \& A, 502, 7}

\bibitem[Sadowski et al.(2011)]{SadwAbra} 
Sadowski A., Abramowicz M., Bursa M. et al., 2011,
\newblock \emph{A\&A, 527, A17}

\bibitem[Salpeter (1964)]{Salp} 
Salpeter, E. E., 1964,
\newblock \emph{ApJ, 140, 796}

\bibitem[Sazonov et al.(2012)]{Sazo}
Sazonov, S., Willner, S. P., Gouldinget, A. D. et al., 2012,
\newblock \emph{ApJ, 757.2, 181}

\bibitem[Schartmann et al.(2005)]{Schart}
Schartmann, M., Meisenheimer, K., Camenzind, M. et al., 2005,
\newblock \emph{A\&A, 437, 861}

\bibitem[Schlegel et al.(1998)]{Schle}
Schlegel, D., J., Finkbeiner, D., P. \& Davis, M., 1998,
\newblock \emph{ApJ, 500, 525}

\bibitem[Schneider et al.(2010)]{Schnei}
Schneider, D. P., Richards, G. T., Hall, P. B. et al., 2010, 
\newblock \emph{ApJ, 139, 2360}

\bibitem[Shakura \& Sunyaev(1973)]{SS}
Shakura, N.~I. \& Sunyaev, R.~A., 1973,
\newblock \emph{AA, 24, 337 (SS)}

\bibitem[Shang et al.(2005)]{Shang05}
Shang, Z., Brotherton, M. S., Green, R. F. et al., 2005,
\newblock \emph{ApJ, 619, 41}

\bibitem[Shen et al.(2011)]{Shenetal11}
Shen, Y., Richards, G. T., Strauss, M. A. et al., 2011,
\newblock \emph{ApJS, 194, 45}

\bibitem[Sikora et al.(2007)]{Siko}
Sikora M., Stawarz L. \& Lesota J.-P., 2007,
\newblock \emph{ApJ, 658, 815}

\bibitem[Simpson(2005)]{Simp}
Simpson, C., 2005,
\newblock \emph{MNRAS, 360, 565-572}

\bibitem[Stalevski et al.(2016)]{Stale}
Stalevski, M., Ricci, C., Ueda, Y. et al., 2016,
\newblock \emph{MNRAS, 458.3, 2288-2302}

\bibitem[Tacconi et al.(1994)]{Tac94}
Tacconi, L. J., Genzel, R., Blietz, M., et al., 1994,
\newblock \emph{ApJ, 426, L77}

\bibitem[Thorne (1974)]{Thorn74}
Thorne, K. S., 1974,
\newblock \emph{ApJ, 191, 507-519}

\bibitem[Toba et al. (2021)]{Toba}
Toba, Y. et al., 2021,
\newblock \emph{ApJ, 912(2), 91}

\bibitem[Trakhtenbrot (2014)]{Trak}
Trakhtenbrot, B., 2014,
\newblock \emph{ApJL, 789(1), L9}

\bibitem[Treister et al.(2008)]{Treister}
Treister E., Krolik J. H. \& Dullemond C., 2008,
\newblock \emph{ApJL, 679, 140-148}

\bibitem[Tristram et al.(2007)]{Trist}
Tristram, K. R. W., Meisenheimer, K., Jaffe, W., et al., 2007,
\newblock \emph{A\&A, 474, 837}

\bibitem[Ueda et al.(2003)]{Ueda}
Ueda Y., Akiyama M., Ohta K. \& Miyaji T., 2003,
\newblock \emph{ApJ, 598, 886-908}

\bibitem[Urry \& Padovani(1995)]{UrPad}
Urry, C. M., \& Padovani, P., 1995,
\newblock \emph{PASP, 107, 803}

\bibitem[Vanden Berk et al.(2001)]{Vanden}
Vanden Berk, D. E.,  Richards, G. T., Bauer, A. et al., 2001,
\newblock \emph{ApJ, 122, 549}

\bibitem[Vasudevan \& Fabian(2009)]{VasuFa}
Vasudevan, R. V. \& Fabian, A. C., 2009,
\newblock \emph{MNRAS, 392, 1124–1140}

\bibitem[Venanzi et al. (2020)]{Vena}
Venanzi, M., H\"{o}nig, S. F. and Williamson, D., 2020,
\newblock \emph{ApJ, 900, 174}

\bibitem[Vestergaard \& Osmer (2009)]{Ves}
Vestergaard, M. \& Osmer, P. S., 2009,
\newblock \emph{ApJ, 699, 800}

\bibitem[Wada(2015)]{Wada}
Wada D., 2015,
\newblock \emph{ApJ, 812, 82}

\bibitem[Wills, Netzer \& Wills(1985)]{Wills}
Wills, B. J., Netzer, H. \& Wills, D., 1985,
\newblock \emph{ApJ, 288, 94}

\bibitem[Wright et al.(2010)]{Wright} 
Wright, E. L.,  Eisenhardt, P. R. M., Mainzer, A. K. et al., 2010, 
\newblock \emph{ApJ, 140, 1868}.

\bibitem[York et al.(2000)]{York}
York, D. G., Adelman, J., Anderson, J. E. et al., 2000, 
\newblock \emph{ApJ, 120, 1579}

\bibitem[Zhao et al. (2021)]{Zhao}
Zhao, X., Marchesi, S., Ajello, M., Cole, D. et al., 2021,
\newblock \emph{A\&A, 650, A57}

\bibitem[Zhuang et al.(2018)]{Zhuang}
Zhuang M. Y., Ho L. C. \& Shangguan J., 2018,
\newblock \emph{ApJ, 862.2, 118}

\bibitem[Zubovas \& King (2019)]{Zubo}
Zubovas, K., \& King, A., R., 2019,
\newblock \emph{arXiv:1908.02630}

\end{thebibliography}
\end{document}